\documentclass[12pt,reqno]{article}

\usepackage[usenames]{color}
\usepackage{amssymb}
\usepackage{amsmath}
\usepackage{amsthm}
\usepackage{amsfonts}
\usepackage{amscd}
\usepackage{graphicx}
\usepackage{tikz}

\usepackage[colorlinks=true,
linkcolor=webgreen,
filecolor=webbrown,
citecolor=webgreen,
hypertexnames=false]{hyperref}

\definecolor{webgreen}{rgb}{0,.5,0}
\definecolor{webbrown}{rgb}{.6,0,0}

\usepackage{color}
\usepackage{fullpage}
\usepackage{float}

\usepackage{psfig}
\usepackage{graphics}
\usepackage{latexsym}
\usepackage{epsf}
\usepackage{breakurl}
\usepackage{multirow}

\newcommand{\seqnum}[1]{\href{https://oeis.org/#1}{\rm \underline{#1}}}

\let\set\mathbb
\def\<#1>{\langle#1\rangle}
\def\i{\mathrm{i}}

\def\O{\operatorname{O}}
\def\sign{\operatorname{sign}}
\def\res{\operatorname{res}}

\begin{document}

\null\vskip 2cm

\theoremstyle{plain}
\newtheorem{theorem}{Theorem}
\newtheorem{corollary}[theorem]{Corollary}
\newtheorem{lemma}[theorem]{Lemma}
\newtheorem{proposition}[theorem]{Proposition}

\theoremstyle{definition}
\newtheorem{definition}[theorem]{Definition}
\newtheorem{example}[theorem]{Example}
\newtheorem{conjecture}[theorem]{Conjecture}

\theoremstyle{remark}
\newtheorem{remark}[theorem]{Remark}

\begin{center}
\vskip 1cm{\LARGE\bf 
    Some D-Finite and Some Possibly D-Finite \\
    \vskip .1in
    Sequences in the OEIS
}
\vskip 1cm
\large
Manuel Kauers\\
Institute for Algebra\\
Johannes Kepler University\\
Altenberger Stra\ss e 69\\
4040 Linz\\
Austria\\
\href{mailto:manuel.kauers@jku.at}{\tt manuel.kauers@jku.at}\\
\ \\
Christoph Koutschan\\
Johann Radon Institute for Computational and Applied Mathematics\\
Austrian Academy of Sciences\\
Altenberger Stra\ss e 69\\
4040 Linz\\
Austria\\
\href{mailto:christoph.koutschan@oeaw.ac.at}{\tt christoph.koutschan@oeaw.ac.at}
\end{center}

\vskip .2in

\begin{abstract}
  In an automatic search, we found conjectural recurrences for some sequences in the
  OEIS that were not previously recognized as being D-finite.
  In some cases, we are able to prove the conjectured recurrence.
  In some cases, we are not able to prove the conjectured recurrence,
  but we can prove that a recurrence exists. 
  In some remaining cases, we do not know where the recurrence might come from. 
\end{abstract}

\section{Introduction}

The {\it On-Line Encyclopedia of Integer Sequences} (OEIS)~\cite{sloane} contains more
than 360,000 sequences of all kinds of different flavors. A prominent flavor is the
class of D-finite sequences, i.e., sequences which satisfy a linear recurrence equation
with polynomial coefficients. Such sequences
are interesting from the point of view of experimental mathematics because extensive
computer algebra support for detecting and proving relations among such sequences is
available. It has been estimated in 2005~\cite{salvy05} and again in 2022~\cite{yurkevich22}
that about 25\% of the sequences in the OEIS fall into this category.

There is a popular technique for searching for potential recurrence equations satisfied
by a sequence of which only the first few terms are known. This technique is known as
``automated guessing'' and is implemented in various computer algebra systems~\cite{salvy94,kauers09a,hebisch11,kauers14b}.
If this method detects a candidate recurrence, it is almost always correct, although
the method does not provide the slightest hint how the relation could be proven. If
the method detects no recurrence, this proves that there is no recurrence of order~$r$
and degree~$d$ for certain $r,d$ such that $(r+2)(d+1)$ is smaller than the number $N$
of available terms. This might mean that the sequence satisfies no recurrence at all,
or that all recurrences it satisfies are too large to be recognized from the available
data.

For the latter situation, we have recently~\cite{kauers22b} introduced a refined variant of
the guessing methodology that is sometimes able to detect recurrences that are beyond
the reach of the classical approach, hereafter referred to as \emph{LA-based guessing} (LA for `linear algebra').
For the present paper, we have scanned the OEIS
for sequences where this new method, hereafter referred to as \emph{LLL-based guessing} (LLL for the lattice reduction
algorithm used within the method), produces interesting output. Applying LLL-based guessing
to all entries of the OEIS where LA-based guessing finds no equation and where at
least 25 and at most 150 terms are available, we detected recurrences in around 600
cases. Going through these cases one by one, many were easily recognized as correct,
and many were easily recognized as wrong, or at least highly implausible. Others were
such that it was easy to compute enough additional terms that LA-based guessing could
find the recurrence.

Here we present the remaining cases, in which we found the guessed recurrence trustworthy
enough to take a closer look at the sequence. An overview is given in Table~\ref{tab:1}.
Using classical techniques, we managed to
prove some of the guessed recurrences, or at least that some recurrence must exist, or
we were able to generate some further terms. These cases are discussed in
Sects.~\ref{sec:tm}--\ref{sec:pa}.
In Sect.~\ref{sec:open}, we list the sequences for which we have found a convincing guess
but no convincing explanation. We invite our readers to take a chance on these sequences.
Results and remarks made in this article have been added to the OEIS entries of the
corresponding sequences. This article is accompanied by a Mathematica notebook
containing all our guessed recurrences, derivations and proofs; it is available at
\url{www.koutschan.de/data/seq/}.

\begin{table}
  \begin{center}
  \def\arraystretch{1.1}
  \begin{tabular}{@{}clllrrrrrl@{}}
    Sect. & Entry & Year & First terms &
    $N$ & $M$ & $L$ & $r$ & $d$ & \\\hline
    \ref{sec:A187990} & \seqnum{A187990} & 2011 & 117,\,181,\,260,\,355,\,467
    & 50 & -- & -- & 1 & 3 & P \rule{0pt}{12pt} \\
    \ref{sec:A177317} & \seqnum{A177317} & 2010 & 1,\,2,\,48,\,2288,\,135040
    & 29 & 60 & 22 & 3 & 14 & P \\
    \ref{sec:A199250} & \seqnum{A199250} & 2011 & 1,\,1,\,14,\,21,\,424,\,571
    & 56 & 98 & 56 & 8 & 18 & P \\
    \ref{sec:A250556} & \seqnum{A250556} & 2014 & 8,\,60,\,302,\,1516,\,7126
    & 47 & 58 & 47 & 9 & 8 & P \\
    \ref{sec:A264947} & \seqnum{A264947} & 2015 & 1,\,60,\,3201,\,184740
    & 20 &  ? &  ? & ? & ? & D \\
    \ref{sec:A265234} & \seqnum{A265234} & 2015 & 1,\,43,\,2592,\,184740
    & 31 & 56 & 27 & 6 & 6 & P \\
    \ref{sec:A172572} & \seqnum{A172572} & 2010 & 90,\,67950,\,90291600
    & 33 & 44 & 17 & 3 & 9 & D \\
    \ref{sec:A172572} & \seqnum{A172671} & 2010 & 90,\,202410,\,747558000
    & 33 & 75 & 25 & 4 & 13 & D \\
    \ref{sec:A188818} & \seqnum{A188818} & 2011 & 2,\,9,\,48,\,256,\,1360
    & 32 & 55 & 26 & 5 & 10 & P \\
    \ref{sec:A306322} & \seqnum{A306322} & 2019 & 1,\,0,\,0,\,25,\,386,\,4657
    & 41 & 63 & 30 & 4 & 14 & P \\
    \ref{sec:A195806} & \seqnum{A195806} & 2011 & 16,\,105,\,496,\,1759,\,5052
    & 32 & 41 & 30 & 4 & 10 & D \\
    \ref{sec:A195806} & \seqnum{A216940} & 2012 & 260,\,27768,\,1664244
    & 37 & 44 & 29 & 1 & 23 & D \\
    \ref{sec:A194478} & \seqnum{A194478} & 2011 & 0,\,0,\,0,\,1,\,337,\,8733
    & 32 & 35 & 19 & 2 & 12 & P \\
    \ref{sec:A215570} & \seqnum{A215570} & 2012 & 1,\,35,\,18720,\,19369350
    & 48 & 68 & 27 & 3 & 15 & O \\
    \ref{sec:A339987} & \seqnum{A339987} & 2020 & 1,\,4,\,90,\,8400,\,1426950
    & 40 & 70 & 24 & 5 & 10 & O \\
    \ref{sec:A269021} & \seqnum{A269021} & 2016 & 1,\,2,\,23,\,588,\,24553
    & 42 &108 & 28 & 4 & 21 & O \\
    \ref{sec:A181198} & \seqnum{A181198} & 2010 & 1,\,1,\,8,\,169,\,6392
    & 27 & 33 & 14 & 2 & 9 & O \\
    \ref{sec:A181198} & \seqnum{A181199} & 2010 & 1,\,1,\,16,\,985,\,141696
    & 26 &103 & 34 & 3 & 24 & O \\
    \ref{sec:A181280} & \seqnum{A181280} & 2010 & 0,\,0,\,0,\,58,\,1629,\,28924
    & 27 & 32 & 26 & 10 & 1 & O \\
    \ref{sec:A253217} & \seqnum{A253217} & 2014 & 0,\,0,\,1,\,19,\,268,\,3568
    & 37 & 53 & 27 & 5 & 9 & O \\
    \ref{sec:A098926} & \seqnum{A098926} & 2004 & 0,\,2,\,12,\,90,\,556,\,5242
    & 34 & 55 & 26 & 8 & 7 & O \\
    \ref{sec:A164735} & \seqnum{A164735} & 2009 & 0,\,0,\,0,\,0,\,0,\,0,\,0,\,1,\,0,\,4
    & 70 & 80 & 66 & 15 & 4 & O \\
  \end{tabular}
  \end{center}
  \caption{%
    $N$ is the number of terms available in the OEIS at the time of writing.\newline
    $M$ is the minimal number of terms that LA-based guessing,
    as implemented in the command \texttt{GuessMinRE} of the package \texttt{Guess.m}~\cite{kauers09a}
    needs in order to detect the recurrence.\newline
    $L$ is the minimal number of terms that LLL-based guessing~\cite{kauers22b} needs in order to
    detect the recurrence.\newline
    $r$~and $d$ are the order and the degree of the recurrence we found.\newline
    In the rightmost column, `P' indicates that the guessed recurrence is proven, `D' means
    that we can prove D-finiteness but not the guessed recurrence, and `O' means that
    the case is open.}
  \label{tab:1}
\end{table}

\subsection{\texorpdfstring{Sequences \seqnum{A237684}}{A237684} and \texorpdfstring{\seqnum{A039836}}{A039836}}\label{sec:A237684}

Conjectures produced by automated guessing can often be trusted, but not always.
Before we get to trustworthy discoveries, let us mention some irregular cases. 

For example, the sequence \seqnum{A237684} is defined as
\[
  a_n = \biggl\lfloor\frac{n\,p(n)}{\sum_{k\leq n} p(k)}\biggr\rfloor,
\]
where $p(n)$ denotes the $n$th prime number. It is known that $p(n)$ is
not D-finite~\cite{flajolet05}, so it may come as a surprise that our LLL-based
guesser finds the astonishingly simple recurrence
\[
  (n-8)a_n + (14-2n)a_{n+1} + (2n-10)a_{n+2} + (4-n)a_{n+3}=0.
\]
To see what is going on here, observe that the first few terms of the
sequence are
\begin{alignat*}1
  &1, 1, 1, 1, 1, 1, 2, 1, 2, 2, 2, 2, 2, 2, 2, 2, 2, 2, 2, 2,\\
  &2, 2, 2, 2, 2, 2, 2, 2, 2, 2, 2, 2, 2, 2, 2, 2, 2, 2, 2, 2,\\
  &2, 2, 2, 2, 2, 2, 2, 2, 2, 2, 2, 2, 2, 2, 2, 2, 2, 2, 2, 2,\\
  &2, 2, 2, 2, 2, 2, 2, 2, 2, 2, 2, 2, 2, 2, 2, 2, 2, 2, 2, 2,\\
  &2, 2, 2, 2, 2, 2, 2, 2, 2, 2, 2, 2, 2, 2, 2, 2, 2, 2, 2, 2.
\end{alignat*}
For at least the next few thousand terms, the sequence continues
with 2's, and the guessed recurrence is correct if and only if the
sequence continues with 2's forever. The guesser did not discover
any interesting pattern but only resonates the obvious observation
that the sequence appears to be ultimately constant. It just chose
the coefficients of the recurrence in such a way that it matches
the finitely many irregular terms in the beginning. Incidentally,
the sequence is indeed constant for $n\geq8$, so after all, the
guessed recurrence happens to be correct; see the recent work of
Axler~\cite{axler19} and the references given in his paper. 

Another example for the same phenomenon is the sequence \seqnum{A039836}, whose
$n$th term is defined as the maximal number $m$ of integers $s_i$ with
$1\leq s_1<s_2<\cdots<s_m\leq n$ such that all sums $s_i+s_j$ with $i\neq j$
are pairwise distinct. The LLL-based guesser finds a recurrence of order~2 and
degree~36 which we do not reproduce here because there is not reason
to believe that it is correct. The initial terms of the sequence are
\begin{alignat*}1
  &1, 2, 3, 3, 4, 4, 4, 5, 5, 5, 5, 5, 6, 6, 6, 6, 6, 6, 7, 7,\\
  &7, 7, 7, 7, 8, 8, 8, 8, 8, 8, 8, 8, 8, 8, 9, 9, 9, 9, 9, 9,\\
  &9, 9, 9, 9, 9, 10, 10, 10, 10, 10, 10, 10, 10, 10, 10, \\
  &10, 10, 11, 11, 11, 11, 11, 11, 11, 11, 11, 11, 11, \\
  &11, 11, 11, 12, 12, 12, 12, 12, 12,
\end{alignat*}
and again, the recurrence only seems to express that the sequence is
constant except at some (finitely many) exceptional indices. This is
not convincing. 

\subsection{\texorpdfstring{Sequence \seqnum{A187990}}{A187990}}\label{sec:A187990}

If a guesser returns a recurrence whose polynomial coefficients encode
that there are some exceptional indices, then it is a good idea to be
skeptical. But we should not be too skeptical either. For example, consider
the sequence \seqnum{A187990}, which counts the number of nondecreasing
arrangements $x_1\leq\dots\leq x_6$ with $x_1,\dots,x_6\in\{-n-4,\dots,n+4\}$ and
$\sum_{i=1}^6 \sign(x_i)\cdot 2^{|x_i|}=0$ where $\sign(0)=1$. LLL-based guessing
delivers the recurrence
\begin{alignat*}1
  &(n-27) (n-26) (n^3+39 n^2+260 n+402) a_{n+1}\\
  &\hbox{\quad\quad} = (n-27) (n-26) (n^3+42 n^2+341 n+702) a_n,
\end{alignat*}
which looks suspicious, because it indicates that $n=27$ is an outlier. We
would probably not expect such an isolated outlier for the sequence, and so
we might be tempted to discard the recurrence as probably wrong.

\begin{table}
\[
  \def\arraystretch{1.2}
  \begin{array}{l|l|c}
    \text{case} & \text{ranges} & \text{number} \\ \hline
    -x_1,-x_2,-x_3,x_3,x_2,x_1 & 1\leq x_1\leq x_2\leq x_3\leq n+4 & \binom{n+6}3
      \rule[-6pt]{0pt}{18pt} \\\hline
    -x_1,-x_2,x_2-1,x_2-1, & \multirow{2}{*}{$1\leq x_2\leq x_1\leq n+4$} &
      \multirow{2}{*}{$\binom{n+5}2$} \\
      \quad x_1-1,x_1-1 & & \\\hline
    -x_1,-x_2,-x_2,x_2+1,  & 1\leq x_1\leq n+4, 1\leq x_2\leq n+3, & \multirow{2}{*}{$(n+3)^2$} \\
      \quad x_1-1,x_1-1 & \quad x_1\neq x_2+1 & \\\hline
    -x_1,-x_1,-x_2,-x_2,x_2+1, & \multirow{2}{*}{$1\leq x_2\leq x_1\leq n+3$} &
      \multirow{2}{*}{$\binom{n+4}2$} \\
      \quad x_1+1 & & \\\hline
    -x_1,-x_2,x_2,x_1-2,x_1-2, & 2\leq x_1\leq n+4, 1\leq x_2\leq n+4, &
      \multirow{2}{*}{$(n+3)^2$} \\
      \quad x_1-1& \quad x_1\neq x_2+1 & \\\hline
    -x_1,-x_1+1,-x_1+1,-x_2, & 2\leq x_1\leq n+3,1\leq x_2\leq n+4, &
      \multirow{2}{*}{$2\binom{n+3}2$} \\
      \quad x_2,x_1+1 & \quad x_1\neq x_2 & \\\hline
    -x_1-3,x_1,x_1,x_1,x_1,x_1+2 & 0\leq x_1\leq n+1 & n+2 \\\hline
    -x_1-3,x_1,x_1,x_1+1, & \multirow{2}{*}{$0\leq x_1\leq n+1$} & \multirow{2}{*}{$n+2$} \\
      \quad x_1+1,x_1+1 & & \\\hline
    -x_1-4,x_1,x_1,x_1+1, & \multirow{2}{*}{$0\leq x_1\leq n$} & \multirow{2}{*}{$n+1$} \\
      \quad x_1+2,x_1+3 & & \\\hline
    -x_1-2,-x_1,-x_1,-x_1, & \multirow{2}{*}{$1\leq x_1\leq n+1$} & \multirow{2}{*}{$n+1$} \\
      \quad -x_1,x_1+3 & & \\\hline
    -x_1-1,-x_1-1,-x_1-1, & \multirow{2}{*}{$1\leq x_1\leq n+1$} & \multirow{2}{*}{$n+1$} \\
      \quad -x_1,-x_1,x_1+3 & & \\\hline
    -x_1-3,-x_1-2,-x_1-1, & \multirow{2}{*}{$1\leq x_1\leq n$} & \multirow{2}{*}{$n$} \\
      \quad -x_1,-x_1,x_1+4 & &
  \end{array}
  \]
  \caption{Case distinction for \seqnum{A187990}.}
  \label{tab:2}
\end{table}

But there is another possible explanation. It could also be that the value $a_{27}$
is incorrect. Indeed, we can derive a closed form for the number of 6-tuples
by case distinction.  In Table~\ref{tab:2} we assume $x_i\geq0$ but not that the
entries appear in the correct order, and in each line we count only those cases
that were not counted in some previous line.

Putting everything together yields $a_n=\frac16(n^3+39 n^2+260 n+402)$ and therefore
$a_{27}=9256$, in contrast to the value $9168$ that was given in OEIS.

\section{Basics about D-finiteness}\label{sec:basics}

We give a quick summary of some basic facts and terminology about D-finite sequences.
Most of this is probably known to most readers, the others are referred to classical
sources~\cite{stanley80,zeilberger90,Zeilberger91,salvy94,bronstein96,petkovsek97,stanley99,kauers10j,Koutschan13a,kauers13,chyzak14}
for further information.
\begin{enumerate}
\item A power series $a(x)=\sum_{n=0}^\infty a_nx^n$ is called \emph{D-finite} if it
  satisfies a linear differential equation with polynomial coefficients, i.e., if
  there are polynomials $p_0,\dots,p_r$, not all zero, such that
  \[
  p_0(x)a(x) + p_1(x)a'(x) + \cdots + p_r(x)a^{(r)}(x) = 0.
  \]
\item A sequence $(a_n)$ is called \emph{D-finite} if it satisfies a linear recurrence
  with polynomial coefficients, i.e., if there are polynomials $p_0,\dots,p_r$, not
  all zero, such that
  \[
  p_0(n)a_n + p_1(n)a_{n+1} + \cdots + p_r(n)a_{n+r} = 0
  \]
  for all $n\in\set N$. Some authors say P-finite or P-recursive instead of D-finite.
\item A sequence $(a_n)$ is D-finite if and only if the corresponding power series $\sum_{n=0}^\infty a_nx^n$
  is D-finite. D-finiteness of sequences and power series is preserved under
  addition and multiplication. If $(a_n)$ and $(b_n)$ are D-finite sequences,
  then so is their interlacing sequence $a_0,b_0,a_1,b_1,a_2,b_2,\dots$.
  If $a(x)$ is D-finite and $b(x)$ is algebraic, then $a(b(x))$ is D-finite.
  All these facts are known as \emph{closure properties} of the class of D-finite
  sequences/series. Closure properties are constructive in the sense that, for
  example, a (provably correct) recurrence for $(a_n+b_n)$ can be computed from
  known recurrences for $(a_n)$ and $(b_n)$.
\item It can be useful to view differential equations and recurrence equations as operators.
  For example, we may write a differential equation in the form
  \[
    (p_0(x) + p_1(x)D + \cdots + p_r(x)D^r)\cdot a(x) = 0,
  \]
  where $D$ denotes the derivation. The operator $p_0(x) + p_1(x)D + \cdots + p_r(x)D^r$
  belongs to a certain non-commutative ring in which the multiplication is defined in such
  a way that it amounts to the composition of operators, i.e., we have $(ML)\cdot a(x)=M\cdot(L\cdot a(x))$
  for any two operators~$M,L$. Note, for example, that we have $Dx=xD+1$ in this ring. 

  An analogous construction is possible for recurrence equations. Instead of the derivation $D$
  we then use the forward shift~$S$, which acts via $S\cdot (a_n)=(a_{n+1})$. In this
  case we have the noncommutativity relation $Sx=(x+1)S$. 
\item 
  If $L$ and $M$ are two operators, we say that $L$ is a \emph{right factor} of $ML$ and
  that $ML$ is a \emph{left multiple} of~$L$. The operator $L$ is called \emph{irreducible}
  if it does not have any nontrivial right factor.
  Note that if $a$ is a solution of an operator~$L$, then it is also a solution of every
  left multiple of~$L$,
  because $L\cdot a=0$ implies $(ML)\cdot a=M\cdot(L\cdot a)=M\cdot0=0$ for every~$M$.
  Conversely, if $a$ is a solution of $ML$, it may or may not be a solution of~$L$, but
  it can be checked algorithmically whether it is.   
\item A bivariate series $a(x,y)$ is called \emph{D-finite} if it is D-finite
  w.r.t.\ $x$ and D-finite w.r.t.~$y$, i.e., if there are polynomials $p_1,\dots,p_r$,
  not all zero, and polynomials $q_1,\dots,q_s$, not all zero, such that
  \begin{alignat*}1
    &p_0(x,y)a(x,y) + p_1(x,y)\frac{d}{dx}a(x,y) + \cdots + p_r(x,y)\frac{d^r}{dx^r}a(x,y) = 0,\\
    &q_0(x,y)a(x,y) + q_1(x,y)\frac{d}{dy}a(x,y) + \cdots + q_s(x,y)\frac{d^s}{dy^s}a(x,y) = 0.
  \end{alignat*}
  The definition extends in the obvious way to series in any (finite) number
  of variables. The definition also applies to series that may involve negative
  or fractional exponents. 
\item Sums and products of multivariate D-finite series are again D-finite (``closure properties'').
  Taking \emph{residues} also preserves D-finiteness. For example, if
  $a(x,y)$ is a bivariate D-finite series, then the series $\res_x a(x,y):=\<x^{-1}>a(x,y)$
  is a univariate D-finite series in~$y$. Also, if we write $a(x,y)=\sum_{n,k} a_{n,k} x^n y^k$,
  then the \emph{diagonal} $(a_{n,n})_{n=0}^\infty$ is a univariate D-finite sequence.
  These operations extend to more variables and they are constructive. 
  Differential equations satisfied by residues or a recurrence equation satisfied
  by the diagonal can be computed by a technique known as \emph{creative telescoping.}
\item Creative telescoping is also used for summation. If $(a_{n,k})$ is a bivariate
  sequence such that its generating function $a(x,y)=\sum_{n,k} a_{n,k}x^ny^k$ is
  D-finite, then the definite sum $\sum_{k=0}^n a_{n,k}$ is a univariate D-finite
  sequence, and we can compute a recurrence for it from a known system of differential
  equations for $a(x,y)$. This applies in particular when $a_{n,k}$ can be written
  as a product of polynomials and binomial coefficients, and it extends to the case
  of more variables and multiple sums.
\end{enumerate}

\section{Transfer matrix method}\label{sec:tm}

\subsection{\texorpdfstring{Sequence \seqnum{A177317}}{A177317}}\label{sec:A177317}

Our first candidate sequence $(a_n)$ counts the number of
permutations of $n$ copies of $\{1,\dots,5\}$ such that any two neighboring
entries differ by at most one. For example, for $n=1$, there are exactly two
such permutations,
\[
  (1,2,3,4,5) \quad\text{and}\quad (5,4,3,2,1),
\]
while for $n=2$ there are more interesting instances, like
\[
  (2, 1, 1, 2, 3, 3, 4, 5, 4, 5) \quad\text{or}\quad (4, 5, 5, 4, 3, 2, 3, 2, 1, 1),
\]
in total $a_2=48$ permutations. From the 29 given terms, the LLL-based guesser finds
a recurrence of order~3 with polynomial coefficients of degree~14,
which roughly looks as follows:
\begin{align*}
  & (n+2)^2 (n+3)^4 \bigl(13113 n^8+\dots+10512\bigr) a_{n+3}\\
  &-2 (n+2)^2 \bigl(668763n^{12}+\dots+20370096\bigr) a_{n+2}\\
  & +(n+1)^2 \bigl(878571n^{12}+\dots+14722560\bigr) a_{n+1}\\
  &-3 n^3 (n+1) (3n+1) (3n+2) \bigl(13113 n^8+\dots+3281160\bigr) a_n = 0.
\end{align*}
The same recurrence can actually be found by using only 22 terms, giving us
some confidence that it is meaningful. In contrast, LA-based guessing requires
at least 60 terms, and therefore could not find it from the available data.

The sequence \seqnum{A177317} is the 5th row of the bivariate sequence
\seqnum{A331562}, whose $i$th row counts the described permutations
with entries in $\{1,\dots,i\}$.
Only the first four rows were already known to be
D-finite. The argument below shows that actually every row is D-finite.

The sequence entries can be computed by dynamic programming, more specifically by
the transfer matrix method~\cite{KramersWannierI,KramersWannierII,stanley99}.
This method is applicable 
whenever the possible choices at a certain position (here: the $k$th position
in the permutation) depend only locally on the previous state (here: the
$(k-1)$st position in the permutation), so that the transition can be modeled
by a finite-state machine. The global condition that each number
must appear exactly $n$ times is taken care of by introducing catalytic
variables: for each~$i$, the variable~$x_i$ records the number of occurrences
of~$i$. Let $p_n\in\set Z[x_1,x_2,x_3,x_4]$ be the permutation-counting
polynomial, whose coefficient of the monomial $x_1^ax_2^bx_3^cx_4^d$ equals
the number of permutations of length~$n$ with $a$ 1's, $b$ 2's,
etc., and $n-a-b-c-d$ 5's, with entries in $\{1,\dots,5\}$ and satisfying
the gap condition. Since we know that the total length is~$n$, we do not need a
variable~$x_5$ to count the 5's. We use the following transfer matrix~$M$,
together with the start vector $v_{\mathrm{init}}$ and the accepting-state
vector $v_{\mathrm{final}}$,
\[
  M = \begin{pmatrix}
    x_1 & x_1 & 0 & 0 & 0 \\
    x_2 & x_2 & x_2 & 0 & 0 \\
    0 & x_3 & x_3 & x_3 & 0 \\
    0 & 0 & x_4 & x_4 & x_4 \\
    0 & 0 & 0 & 1 & 1 \\
  \end{pmatrix},\quad
  v_{\mathrm{init}} = (x_1,x_2,x_3,x_4,1),\quad
  v_{\mathrm{final}} = \begin{pmatrix} 1 \\ 1 \\ 1 \\ 1 \\ 1\end{pmatrix}
\]
to express the permutation-counting polynomial as a matrix-vector product:
\[
  p_n(x_1,x_2,x_3,x_4) = v_{\mathrm{init}}\cdot M^{n-1}\cdot v_{\mathrm{final}}.
\]
Now, the sequence entries can be obtained by a simple coefficient extraction:
\[
  a_n = \bigl\langle x_1^nx_2^nx_3^nx_4^n\bigr\rangle\, p_{5n}(x_1,x_2,x_3,x_4)
  = \bigl\langle x_1^nx_2^nx_3^nx_4^n\bigr\rangle
  \bigl(v_{\mathrm{init}}\cdot M^{5n-1}\cdot v_{\mathrm{final}}\bigr).
\]
Although the matrix is of small size, and despite the fact that we have
already saved one variable, it is quite time-consuming to compute the values
$a_n$ in this way, because the four-variable polynomials grow very rapidly.
For example, computing $a_{12}$ takes about four minutes and produces a vector
of more than one gigabyte in size.

The method could be optimized, e.g., by truncating the intermediate polynomials and
omitting all terms with exponents greater than~$n$. However, instead of using the
transfer matrix method to compute~$a_n$ for specific values of~$n$, it is more interesting to
employ it for deriving a closed form for the five-variable generating function
$F(x_1,x_2,x_3,x_4,t) = \sum_{n=0}^\infty p_n(x_1,x_2,x_3,x_4)t^n$.

For this purpose, recall the explicit formula~\cite[Thm.~4.7.2]{stanley99} for
the generating function of the sequence appearing in the $(i,j)$th entry of a
matrix power~$M^n$
\begin{equation}\label{eq:matpowergf}
  \sum_{n=0}^\infty \bigl(M^n\bigr)_{i,j}\cdot t^n =
  (-1)^{i+j} \,\frac{\det(I_\ell - t\,M)^{[j,i]}}{\det(I_\ell - t\,M)},
\end{equation}
where the exponent $[j,i]$ indicates the removal of the $j$th row and the
$i$th column of the matrix $I_\ell-tM$. Hence, the generating function~$F$
is just a certain linear combination of such rational functions, determined
by the vectors $v_{\mathrm{init}}$ and $v_{\mathrm{final}}$. An explicit computation
gives
\begin{multline*}
  F(x_1,x_2,x_3,x_4,t) = \Bigl(
  2 t^3 x_3 (x_1 x_2+x_1 x_4 x_2 +x_4 x_2+x_1 x_4) - t^2 x_3 (x_2 x_1-3x_1\\
  +x_2 x_4+x_4) - 2t (x_3 x_1+x_4 x_1+x_1+x_2+x_3+x_2 x_4)+x_1+x_2+x_3+x_4+1\Bigr)\\
  \Big/ \Bigl( - t^4 x_3 (x_1 x_2+x_1 x_4 x_2+x_4 x_2+x_1 x_4) \\
  + t^3 x_3 (x_2 x_1-x_1+x_2 x_4+x_4) + t^2 (x_3 x_1+x_4 x_1+x_1+x_2+x_3\\
  +x_2 x_4)- t (x_1+x_2+x_3+x_4+1)+1 \Bigr).
\end{multline*}
Using this generating function, the sequence terms can be expressed as a residue,
\[
  a_n = \bigl\langle x_1^nx_2^nx_3^nx_4^nt^{5n-1}\bigr\rangle F(x_1,x_2,x_3,x_4,t) =
  \res_{x_1,x_2,x_3,x_4,t} \frac{F(x_1,x_2,x_3,x_4,t)}{(x_1x_2x_3x_4)^{n+1}\,t^{5n}}.
\]
A recurrence equation for the residue can be derived by creative telescoping.
Here, we have to apply it five times, once for each
variable, which takes about 10 minutes in total, using \texttt{HolonomicFunctions.m}~\cite{koutschan10c}.
The result is exactly the guessed order-3 recurrence, which proves that
the guess was indeed correct.

\begin{theorem} \seqnum{A177317} is D-finite and satisfies a recurrence of
  order~3 and degree~14.
\end{theorem}

\subsection{\texorpdfstring{Sequence \seqnum{A199250}}{A199250}}\label{sec:A199250}

The next sequence deals with a similar counting problem, but now for
two-dimensional arrangements. Its description in the OEIS reads as follows:
``number of $n\times2$ arrays with values $\{0,\dots,3\}$ introduced in row major
order, the number of instances of each value within one of each other,
and no element equal to any horizontal or vertical neighbor.''

Using the 56 terms given in the OEIS, a linear recurrence of order~22 and
coefficient degree~3 can be guessed. We realize that this is not the minimal
one: when more terms are used (they can conjecturally be produced, e.g., by
applying the guessed order-22 recurrence), then a recurrence of order~8 and degree~18 can
be found, which happens to be a right factor of the previous one, when viewed as
operators. It is very unlikely that an artifact recurrence has such a right
factor, and thus our guess appears to be trustworthy.

Also this sequence can be computed with the transfer matrix method. Since
horizontal neighbors must be different, there are 12 possible rows that can
appear in such arrays,
\[
  (0, 1), (0, 2), (0, 3), (1, 0), (1, 2), (1, 3),
  (2, 0), (2, 1), (2, 3), (3, 0), (3, 1), (3, 2),
\]
each of which represents a state. The condition that vertical neighbors must
be unequal determines a finite-state machine that encodes which rows can
potentially follow any given row. As in the previous section, one introduces
catalytic variables to implement the global condition that each number must
appear equally often in the array (resp., ``almost equally often'' if the
number of rows is odd). This yields the following $12\times12$-matrix~$M$:
\[
  M =
  \left(\begin{array}{cccccccccccc}
    0 & 0 & 0 & x y & y z & y & x z & 0 & z & x & 0 & z \\
    0 & 0 & 0 & x y & 0 & y & x z & y z & z & x & y & 0 \\
    0 & 0 & 0 & x y & y z & 0 & x z & y z & 0 & x & y & z \\
    x y & x z & x & 0 & 0 & 0 & 0 & y z & z & 0 & y & z \\
    x y & 0 & x & 0 & 0 & 0 & x z & y z & z & x & y & 0 \\
    x y & x z & 0 & 0 & 0 & 0 & x z & y z & 0 & x & y & z \\
    x y & x z & x & 0 & y z & y & 0 & 0 & 0 & 0 & y & z \\
    0 & x z & x & x y & y z & y & 0 & 0 & 0 & x & 0 & z \\
    x y & x z & 0 & x y & y z & 0 & 0 & 0 & 0 & x & y & z \\
    x y & x z & x & 0 & y z & y & 0 & y z & z & 0 & 0 & 0 \\
    0 & x z & x & x y & y z & y & x z & 0 & z & 0 & 0 & 0 \\
    x y & 0 & x & x y & 0 & y & x z & y z & z & 0 & 0 & 0
  \end{array}\right).
\]
Its $(i,j)$-entry equals~$0$ if state~$i$ and state~$j$ agree on their first or second
position. Otherwise the $(i,j)$-entry of~$M$ equals $x^ay^bz^c$, where $a$
(or $b$ or $c$, resp.) counts the number of $0$'s (or $1$'s or $2$'s, resp.)
in state~$j$. The condition that numbers are introduced in row-major order
forces the first row to be $(0,1)$, so this is the only initial state, while
all states can be accepting states, and thus we define
\begin{alignat*}1
  v_{\mathrm{init}} &= (xy,0,0,0,0,0,0,0,0,0,0,0), \\
  v_{\mathrm{final}} &= (1,1,1,1,1,1,1,1,1,1,1,1)^\top.
\end{alignat*}
Then, for each $n\geq1$, the polynomial
$p_n(x,y,z) = v_{\mathrm{init}}\cdot M^{n-1}\cdot v_{\mathrm{final}}$
counts the number of such arrays, disregarding the balancing of the number of
occurrences of $0$'s, $1$'s, $2$'s, and $3$'s. Hence, we are interested in
the coefficient of $(xyz)^{n/2}$ in $p_n(x,y,z)$ if $n$ is even, or in the sum
of the six coefficients of $(xy)^{(n-1)/2}z^{(n+1)/2}$, \dots, $(xy)^{(n+1)/2}z^{(n-1)/2}$,
\dots, if~$n$ is odd. Finally, this number has to be divided by~$2$,
in order to discard all solutions where a $3$ is introduced before a $2$ (in
row-major order).

With this method it takes less than half an hour to compute the first $100$
terms of the sequence, allowing us to cross-check our conjecture with terms that were
not used for the guessing. Moreover, the transfer-matrix construction implies that
the sequence is D-finite, and it enables us to deduce a provably correct recurrence.
For the generating function of the full counting sequence,
$F(x,y,z,t)=\sum_{n=0}^{\infty} p_n(x,y,z)t^n$, several applications of
\eqref{eq:matpowergf} yield the following closed form:
\begin{alignat*}1
  F(x,y,z,t) &=
  \frac{txy \sum_{i=1}^{12} (-1)^{i+1}\,\det(I_{12}-tM)^{[1,i]}}{\det(I_{12}-tM)}\\
  &=
  \frac{t x y (t z+1)}{1 - t x - t y - t x y - t z - t x z - t y z - 7 t^2 x y z}.
\end{alignat*}
The desired recurrence can now be obtained via creative telescoping.
For example, for even~$n$, we compute a recurrence for
\[
  \res_{x,y,z,t} \frac{1}{xyzt}\frac{F(x,y,z,t)}{(xyz)^n\,t^{2n}}.
\]
The result, which is an order-6 and degree-17 recurrence for $(a_{2n})$, is obtained in
about a minute. Slightly more complicated is the case of odd~$n$, for which
we deduce a recurrence of order~6 and degree~22. Both are not minimal-order,
but combining them results in a recurrence of order~24 and degree~79 for~$(a_n)$.
The latter is a left multiple of the guessed recurrence operator, therefore
allowing us to prove that the guess is correct.

\begin{theorem}
  \seqnum{A199250} is D-finite and satisfies a recurrence of order~8
  and degree~18. The subsequence formed by the even (resp., odd) indices
  satisfies a recurrence of order~4 and degree~8 (resp.,~10).
\end{theorem}

\subsection{\texorpdfstring{Sequence \seqnum{A250556}}{A250556}}\label{sec:A250556}

It is not always easy to see whether the transfer matrix method can be applied,
and if so, what is a suitable set of states. Consider for example the sequence
\seqnum{A250556}, which is defined as
\[
  a_n := \bigl|\bigl\{v\in\{0,1,2,3\}^{n+2} \mathrel{\big|}
  \exists\,s\in\{-1,+1\}^n: \Delta^2(v)\cdot s=0 \bigr\}\bigr|,
\]
where $\Delta(v_1,\dots,v_n):=(v_2-v_1,\dots,v_n-v_{n-1})$ is the forward
difference operator.  It is not completely obvious how to translate the
conditions on the arrays~$v$ into states, because we have to consider all
possible sign vectors for combining their second differences to~$0$. To
address this problem, we introduce states that encode the following
information:
\begin{enumerate}
\item The last two entries of the array, since they are needed to compute the second
  difference when appending another entry to the array.
\item The set of numbers that can be produced by taking the scalar product of
  the second differences with all possible sign vectors.
\end{enumerate}
Note that for the second item, it suffices to store only the absolute values of
these numbers, since the corresponding negative numbers could be produced by switching
all signs in the sign vector.

For example, consider the state $(3,1,\{1,5\})$, which means that the array that
was produced so far is of the form $(\dots,3,1)$ and that all signed sums
of its second differences sum up to either $1$ or~$5$ (or, of course, to $-1$
or $-5$).  We wish to extend the array by a~$1$. The new second difference
that we can build is $3-2\cdot1+1=2$. Hence we add or subtract~$2$ to each
number in the list, yielding the new state $(1,1,\{1,3,7\})$. Note that
$1-2=-1$ has turned into a $+1$ by our nonnegativity convention.

The problem is that the signed sums of the second differences can get
arbitrarily large as the arrays get longer. Of course, if we bound the number
of sequence terms we wish to compute, then we could devise an upper bound for
these signed sums. Then with a fixed transfer matrix we could compute a certain
finite number of sequence terms. Fortunately, we can do better:
we derive a global upper bound~$B$ and show that it is sufficient to
store only signed sums up to~$B$, independent on the length of the arrays. This
bound~$B$ must have the property that for any sequence of signed second
differences that add up to~$0$ and whose partial sums exceed~$B$, there must
exist another sign vector, that combines these second differences to~$0$
without exceeding~$B$. Here is an example showing that $B\geq19$: the array
\[
  (1, 3, 0, 2, 0, 3, 0, 3, 0, 3, 1)
\]
has the second differences
\[
  (-5, 5, -4, 5, -6, 6, -6, 6, -5)
\]
which combine to~$0$ using the sign vector
\[
  (-1, 1, -1, 1, 1, -1, 1, -1, -1)
\]
(or its additive inverse). Note that there are no other sign vectors that
produce~$0$. The partial sums in the signed sum $5+5+4+5-6-6-6-6+5=0$
go all the way up to~$19$ before they finally descend to~$0$.

We argue that actually $B=19$, i.e., that there is no example like the one
above where the partial sums are forced to exceed~$19$. For this purpose, we
have to identify all pairs $(S_1,S_2)$ of multisets with values in
$\{1,\dots,6\}$ such that $\sum(S_1)=\sum(S_2)>19$, but such that there are no
nontrivial subsets $T_1\subset S_1$ and $T_2\subset S_2$ with
$\sum(T_1)=\sum(T_2)$. Hence, the only way that a signed sum of $S_1\cup S_2$
equals~$0$ is that all elements in $S_1$ have the same sign, and all elements
in $S_2$ have the opposite sign. Here are all possible choices for $S_1$ and~$S_2$:
\begin{alignat*}{2}
  S_1 &= \{1, 1, 6, 6, 6\},&\quad S_2 &= \{5, 5, 5, 5\}, \text{ or} \\
  S_1 &= \{2, 6, 6, 6\},& S_2 &= \{5, 5, 5, 5\}, \text{ or} \\
  S_1 &= \{5, 5, 5, 5\},& S_2 &= \{4, 4, 4, 4, 4\}, \text{ or} \\
  S_1 &= \{6, 6, 6, 6\},& S_2 &= \{4, 5, 5, 5, 5\}, \text{ or} \\
  S_1 &= \{6, 6, 6, 6, 6\},& S_2 &= \{5, 5, 5, 5, 5, 5\}.
\end{alignat*}
The first two possibilities can be excluded, because in the array of second
differences a $\pm6$ can never be followed by a $\pm1$ or by a $\pm2$.
For the remaining three possibilities, we can do an exhaustive search:
build all permutations of $S_1\cup (-S_2)$ that have a partial sum $>19$,
for each of them apply suitable sign vectors (it is easy to see that an
array of second differences with values in $\{\pm4,\pm5,\pm6\}$ must have
alternating signs), and then construct all corresponding arrays~$v$. The
final outcome is that there are no such arrays~$v$, proving that $B=19$
is the desired bound.

Next, a suitable set of states has to be defined. Naively, one could expect
that 16,777,200 states are necessary, since there are 16 possibilities
for the last two entries of the array and $2^{20}-1$ nontrivial subsets of
$\{0,\dots,19\}$. A closer inspection reveals that we can work with much fewer
states. From the transition rule between the states it is apparent that the
reachable numbers in each state are either all even or all odd. Hence it
suffices to take all nontrivial subsets of $\{1,3,5,\dots,19\}$ and of
$\{0,2,4,\dots,18\}$, yielding $16\cdot(2^{10}-1)\cdot2=32{,}736$ states.
Still, this set contains many unreachable states, for example when the set
of possible signed sums has a gap greater than~$12$.
Eliminating all such useless states results in a set of 2484 states.

Using the corresponding $2484\times2484$ transfer matrix, which contains only
$0$'s and $1$'s, one can easily compute hundreds or thousands of sequence
terms in almost no time (0.6s for the first 1000 terms, for example).  The
matrix formulation also implies directly that the sequence is D-finite. Since there are no
catalytic variables, we can directly derive a rational function expression for
its generating function. Hence, the sequence is even C-finite, i.e., it
satisfies a linear recurrence with constant coefficients.  The start vector
$v_{\mathrm{init}}$ has 60 nonzero entries and the accepting-state vector
$v_{\mathrm{final}}$ has~720 nonzeros. Instead of applying the determinant
formula~\eqref{eq:matpowergf} $60\cdot720=43{,}200$ times (each case taking
about three seconds), we compute the signed sum of all $(i,j)$-minors, where
$j$ is a fixed nonzero position in $v_{\mathrm{init}}$ and $i$ runs through
all nonzero positions of $v_{\mathrm{final}}$, by taking the determinant of
the matrix $I_\ell-t\,M$ with the $j$th column being replaced by
$v_{\mathrm{final}}$ (for each $j$ this takes about 30 seconds). Putting
everything together, we obtain the generating function
\begin{align*}
  &{-2 t} \bigl(32 t^{27}-56 t^{26}+508 t^{25}-300 t^{24}+684 t^{23}-1296 t^{22}-1324 t^{21} \\
  &\qquad{}-202 t^{20}+403 t^{19}+4173 t^{18}+1985 t^{17}+903 t^{16}-4504 t^{15} \\
  &\qquad{}-4178 t^{14}-3614 t^{13}+1666 t^{12}+2087 t^{11}+3597 t^{10}+406 t^9 \\
  &\qquad{}+38 t^8-1231 t^7-453 t^6-139 t^5+115 t^4+73 t^3-3 t^2+2 t+4\bigr) \\
  &\big/ \bigl((t-1)^3 (t+1)^2 (2 t-1) (4 t-1) (t^2+1)^2 (2 t^3-1)^2 \bigr).
\end{align*}
The C-finite recurrence for \seqnum{A250556} can be read off from its denominator:
\begin{align*}
  & a_{n+17} - 7 a_{n+16} + 14 a_{n+15} - 12 a_{n+14} + 26 a_{n+13} - 42 a_{n+12}\\
  &{}+ 8 a_{n+11} - 4 a_{n+10} + a_{n+9} + 73 a_{n+8} - 58 a_{n+7} + 44 a_{n+6} \\
  &{}- 84 a_{n+5} + 8 a_{n+4} + 36 a_{n+3} - 28 a_{n+2} + 56 a_{n+1} - 32 a_n = 0.
\end{align*}
This recurrence can be found with guessing from $a_{12},\dots,a_{47}$; the first values
$a_1,\dots,a_{11}$ are exceptional and do not satisfy this recurrence (note that
the numerator degree exceeds the denominator degree by~$10$). Without
this additional knowledge it is not possible to find anything with classical
linear algebra guessing. In contrast, the LLL-based guesser finds a recurrence
of order~22 and degree~1, which is a right factor of the order-27 operator.
The minimal recurrence however is of order~9 and degree~8.

\begin{theorem}
  \seqnum{A250556} is D-finite and satisfies a recurrence of order~9 and degree~8.
\end{theorem}

\subsection{\texorpdfstring{Sequence \seqnum{A264947}}{A264947}}
\label{sec:A264947}

Even for innocent-looking sequences it can sometimes be very hard to compute
their terms and find a recurrence. \seqnum{A264947} enumerates $4\times n$
arrays containing $n$ copies of $\{0,1,2,3\}$ with no equal horizontal neighbors.
(Moreover, new values in the array should be introduced sequentially from~$0$,
but this condition is not so relevant, as it just divides the final count by
$4!=24$.)

The OEIS lists only 20 terms. Can we compute more, and/or derive a
recurrence equation, since this problem is an obvious application of the
transfer matrix method? From what we have seen in the previous sections, it
is clear that the states are the $4^4=256$ possible columns and that we have
to introduce three catalytic variables $x,y,z$ to count the occurrences of
$0,1,2$, respectively. Therefore, we know for sure that \seqnum{A264947} is
D-finite.

However, things are computationally expensive, because the matrix has
considerable dimensions ($256\times256$) and because it contains three
variables. With quite some effort we were able to compute 80 terms of the
sequence. After about one month of non-parallelized computation our compute
server with 256\,GB ran out of memory. Unfortunately, the data obtained before
the crash is still not enough to guess a recurrence (which we know for sure must exist). To get an
idea of the difficulty of this problem, compare with the simpler case of
$3\times n$ arrays with $n$ copies of $\{0,1,2\}$ (\seqnum{A264946}): here
the recurrence has order~9 and degree~13, and we need 63 terms to find
it with LLL-based guessing. The 104 terms given in the OEIS are just
sufficient to find the recurrence with LA-based guessing (and this is
why it did not make it into our collection).

Likewise, we did not succeed to compute the rational function expression for
the four-variable generating function: computing the determinants appearing
in~\eqref{eq:matpowergf} turned out to be prohibitively expensive. We tried
to compute one of the $256$ determinants, but aborted the computation after
five days.

\begin{theorem}
  \seqnum{A264947} is D-finite.
\end{theorem}

It remains an open problem to find a provably correct recurrence for the
sequence~\seqnum{A264947}.

\section{Lattice walks}\label{sec:walks}

\subsection{\texorpdfstring{Sequence \seqnum{A265234}}{A265234}}
\label{sec:A265234}

Changing a small detail can sometimes make a big difference. For example, if
we change the condition ``no equal horizontal neighbors'' in \seqnum{A264947}
from the previous section into ``no equal vertical neighbors'', then the
problem becomes significantly simpler.

This time, the condition on neighbors can be satisfied by making a suitable
selection of admissible columns---there are 108 which do not have
equal neighbors. There are no further restrictions concerning which column can
follow another one. In principle, one could again model this process by a
transfer matrix, but it is more efficient to take a slightly different
viewpoint. Consider the integer lattice $\set Z^3$ and interpret the point
$(x,y,z)$ as having seen $x$ 0's, $y$ 1's, and $z$ 2's,
when filling the array from left to right. Adding a column to the array then
corresponds to making a step in this lattice. Note that different columns may
correspond to the same step: for example, $(1,0,1,3)^\top$ and
$(3,1,1,0)^\top$ both correspond to the step $(1,2,0)$. In this
interpretation, the $n$th sequence term counts the number of walks of
length~$n$, starting at the origin $(0,0,0)$ and ending at $(n,n,n)$. By
construction, these walks will never leave the first octant, and hence,
sequence \seqnum{A265234} can be viewed as an unrestricted walk enumeration
problem in 3D. Using the set of admissible columns, we define the
stepset polynomial
\begin{align*}
  s(x,y,z) ={} & 2 x^2 + 6 x y + 6 x^2 y + 2 y^2 + 6 x y^2 + 2 x^2 y^2 + 6 x z \\
  &{}+ 6 x^2 z + 6 y z + 24 x y z + 6 x^2 y z + 6 y^2 z + 6 x y^2 z \\
  &{}+ 2 z^2 + 6 x z^2 + 2 x^2 z^2 + 6 y z^2 + 6 x y z^2 + 2 y^2 z^2.
\end{align*}
The generating function of \seqnum{A265234} can then be obtained as the
diagonal of the rational function
\[
  \frac{1}{1-t\, s(x,y,z)},
\]
divided by $24$ to account for permutations of the numbers $0,1,2,3$. Creative
telescoping delivers exactly the guessed order-6 recurrence, taking less than
a minute. The sequence terms could also be computed via
\[
  a_n = \frac{1}{24}\bigl\langle x^ny^nz^n \bigr\rangle \bigl(s(x,y,z)\bigr)^n,
\]
which takes about 100s for 56 terms (this is the amount of data necessary for
LA-based guessing).
\begin{theorem}
  \seqnum{A265234} is D-finite and satisfies a recurrence of order~6 and degree~6.
\end{theorem}

\subsection{\texorpdfstring{Sequence \seqnum{A172572} and \seqnum{A172671}}{A172572 and A172671}}
\label{sec:A172572}

These two sequences count the number of $\{0,1\}$-arrays or $\{0,1,2\}$-arrays,
respectively, of dimension $3n\times6$ with row sums~$2$ and column sums~$n$.
Hence, for \seqnum{A172572} the row-sum condition yields exactly $\binom62=15$
possibilities for what a row can look like:
\[
  R_1 = (1,1,0,0,0,0), \;
  R_2 = (1,0,1,0,0,0), \dots, \;
  R_{15} = (0,0,0,0,1,1).
\]
Let $c_i$ denote the number of occurrences of $R_i$ in the final array.
The condition on the column sums translates into
\[
  \sum_{i=1}^{15} c_i R_i = (n,n,n,n,n,n),
\]
which yields six linear equations for the~$c_i$. Their general solution is
\begin{align*}
  c_5 &= n-c_1-c_2-c_3-c_4, \\
  c_9 &= n-c_1-c_6-c_7-c_8, \\
  c_{12} &= n-c_2-c_6-c_{10}-c_{11}, \\
  c_{13} &= 2n-c_1-c_2-c_3-c_4-c_6-c_7-c_8-c_{10}-c_{11}, \\
  c_{14} &= c_1+c_2+c_4+c_6+c_8+c_{11}-n, \\
  c_{15} &= c_1+c_2+c_3+c_6+c_7+c_{10}-n.
\end{align*}
Note that the condition on the number of rows, $\sum_{i=1}^{15} c_i=3n$, is a
consequence of these equations. For each admissible choice of the~$c_i$,
the number of arrays that can be built by permuting the corresponding numbers
of rows is given by the multinomial coefficient
\[
  \binom{3n}{c_1, \dots, c_{15}}.
\]
The total number of arrays~$a_n$ is then obtained by summing over the nine
remaining free variables among the~$c_i$, and by replacing the other ones
by the linear expressions displayed above:
\[
  a_n = \sum_{c_1,c_2,c_3,c_4,c_6,c_7,c_8,c_{10},c_{11}}
  \binom{3n}{c_1,c_2,c_3,c_4,n-c_1-c_2-c_3-c_4,c_6, \dots}.
\]
We have omitted the summation ranges here, since the sum has natural
boundaries. Instead, one could fix the range $0\leq c_i\leq n$ for each
variable, or even more refined summation ranges, implied by the condition that
all lower entries of the multinomial coefficient must be nonnegative and at
most~$3n$. This nine-fold sum can be reduced by means of the Chu-Vandermonde
identity
\[
  \sum_{k=0}^r \binom{m}{k} \binom{n}{r-k} = \binom{m+n}{r}.
\]
Instantiating it with $k=c_{11}$, $r=n-c_2-c_6-c_{10}$,
$m=2n-c_1-c_2-c_3-c_4-c_6-c_7-c_8-c_{10}$, and $n=c_1+c_4+c_8-c_{10}$, we can
eliminate the last summation. This can be done similarly for the summations
w.r.t.~$c_8$ and~$c_4$, so that we obtain the following six-fold sum:
\begin{multline*}
  a_n = \sum_{c_1=0}^n \sum_{c_2=0}^{n-c_1} 
  \sum_{c_3=0}^{n-c_1-c_2} 
  \sum_{c_6=0}^{\min\{n-c_1,n-c_2\}} 
  \sum_{c_7=0}^{\min\{n-c_1-c_6,n-c_3\}}
  \sum_{c_{10}=\max\{0,n-c_1-c_2-c_3-c_6-c_7\}}^{\min\{n-c_2-c_6,n-c_3-c_7\}}
  \\
  \Bigl((3n)! \, (4n-2c_1-2c_2-2c_3-2c_6-2c_7-2c_{10})! \Bigr) \Big/ \Bigl(
  c_1! \, c_2! \, c_3! \, c_6! \, c_7! \, c_{10}! \\
  (n-c_1-c_2-c_3)! \, (n-c_1-c_6-c_7)! (n-c_2-c_6-c_{10})! \, (n-c_3-c_7-c_{10})! \\
  \bigl((2n-c_1-c_2-c_3-c_6-c_7-c_{10})!\bigr)^2 \,
  (c_1+c_2+c_3+c_6+c_7+c_{10}-n)!
  \Bigr)
\end{multline*}
At this point it is clear that \seqnum{A172572} is D-finite. 
However, deriving a recurrence from this sum representation via creative
telescoping is still a challenging task. We were not able to complete it
in reasonable time.

Instead, one can use this formula to compute some further terms of the sequence.
Implementing it in Mathematica, and taking into account some of
the symmetries that follow from permuting the columns of the array,
we get the following timings for computing the $n$th term of the sequence:
\[
  \begin{array}{r|rrrrrcrrrcr}
     n & 12 & 14 & 16 & 18 & 20 & \cdots & 28 & 30 & 32 & \cdots & 44
     \\ \hline
     \text{time (s)}\rule{0pt}{10pt} &
     0.49 & 1.02 & 1.91 & 3.53 & 6.09 & \cdots & 40.1 & 57.1 & 81.6 & \cdots & 518 \\
  \end{array}
\]
The computation of the first 33 terms that were given in OEIS took 549s in
total, while the computation time for the first 44 terms that are needed for
LA-based guessing was 3566s. Note also that the above formula allows one to
compute the $n$th term of the sequence, without computing all the previous ones.

Alternatively, the $\{0,1\}$-arrays counted by \seqnum{A172572} can be
interpreted as walks in the first orthant $\set N^6$ of the six-dimensional
integer lattice, starting at the origin, and with allowed step set
${\cal S}=\{R_1,\dots,R_{15}\}$. The column sum condition implies that we are
interested in the number of walks that end on the diagonal point $(n,n,n,n,n,n)$.
To determine this number, we generate a six-dimensional array~$A$, such that
the entry $A_{n_1,n_2,n_3,n_4,n_5,n_6}$ records the number of walks ending at
position $(n_1,n_2,n_3,n_4,n_5,n_6)$ and using only steps from~$\cal S$.  The
entries of this array can be computed by means of the multivariate C-finite
stepset recurrence
\begin{equation}\label{eq:steprec}
  A_{n_1,n_2,n_3,n_4,n_5,n_6} = \sum_{s\in{\cal S}}
  A_{n_1-s_1,n_2-s_2,n_3-s_3,n_4-s_4,n_5-s_5,n_6-s_6},
\end{equation}
with the initial condition $A_{0,0,0,0,0,0}=1$ and the boundary condition that
$A_{n_1,n_2,n_3,n_4,n_5,n_6}=0$ whenever at least one of the six indices $n_i$ is
negative.  Note that each walk ending at $(n_1,n_2,n_3,n_4,n_5,n_6)$
consists of exactly $(n_1+n_2+n_3+n_4+n_5+n_6)/2$ steps, and thus the
length of the walks does not need to be recorded separately. Several optimizations
can make this enumeration more time- and memory-efficient. First,
we exploit the symmetry that follows from permuting the columns of
the $\{0,1\}$-array, i.e., the coordinates of the array~$A$, which means that
it suffices to record only values for $n_1\geq n_2\geq \dots\geq n_6$.
Second, since for computing the walks with $k$ steps one only
needs the information about walks with $k-1$ steps, we can discard the data
related to shorter walks, which has the effect that only a five-dimensional
array has to be kept in memory. Of course, whenever $k$ is divisible by~$3$,
the diagonal entry should be saved, as it contains the sequence term~$a_{k/3}$.
If one aims at computing $a_1,\dots,a_n$ for prescribed fixed~$n$, then one can confine the array to
$\{0,1,\dots,n\}^5$, because walks that have left this hypercube can never
come back to a diagonal position inside the hypercube. With this approach, we
obtained the first $33$ terms in 232s, while the $44$ terms that are required
for LA-based guessing took 1053s.

We see that this procedure is faster than the previous one, at least when one
wants to compute all terms of the sequence up to a certain index. The
disadvantage is that extending the sequence requires a complete restart of
the computation (or one has to omit some of the optimizations described
above).

In any case, the walk viewpoint allows us to express the generating function of
the sequence \seqnum{A172572} as the diagonal of a six-variable rational function whose
denominator is the stepset polynomial, given as the characteristic polynomial
of the recurrence~\eqref{eq:steprec},
\[
  \sum_{n=0}^\infty a_n x^n = \operatorname{diag}
  \frac{1}{1-x_1x_2-x_1x_3-x_1x_4-\dots-x_4x_5-x_4x_6-x_5x_6}.
\]
From this representation it again follows immediately that the generating
function is D-finite. A recurrence for $(a_n)$ can in principle be derived by
applying creative telescoping to the corresponding six-fold integral, but
similar to the six-fold sum before, we did not manage to complete this task in
reasonable time (the computation was aborted after one month).  We therefore
propose our guessed recurrence as a conjecture to the reader, which we present
in compact form by dividing out a hypergeometric factor.

\begin{conjecture}
  If $(a_n)$ denotes the sequence \seqnum{A172572} then for 
  $\tilde{a}_n := \frac1{\binom{3n}n}a_n$ we have
  \begin{align*}
    & (n+3)^4 (62 n^2+217 n+191) \tilde{a}_{n+3} \\
    & {}-6 (5084 n^6+68634 n^5+383756 n^4+1137319 n^3\\
    &\qquad +1884032 n^2+1653960 n+601185) \tilde{a}_{n+2} \\
    & {}-4 (2 n+3) (31372 n^5+313720 n^4+1227805 n^3\\
    &\qquad+2354425 n^2+2220988 n+827860) \tilde{a}_{n+1} \\
    & {}+6000 (n+1)^2 (2 n+1) (2 n+3) (62 n^2+341 n+470) \tilde{a}_n = 0.
  \end{align*}
\end{conjecture}

The sequence \seqnum{A172671} is very similar, the only difference being that now also
2's are allowed as entries in the array. This increases the number of possible rows
to $\binom62+\binom61=21$. Performing a similar analysis as for \seqnum{A172572},
we find an eleven-fold hypergeometric sum representation, which however is not useful
for any practical purposes.
Here, it is much better to treat the corresponding walk counting problem,
which is still in the six-dimensional integer lattice, but now with a stepset of
size~21. Again, we only succeeded to compute some more sequence terms (the already
available terms $a_1,\dots,a_{33}$ took 298s, while $a_1,\dots,a_{75}$ that are needed
for LA-based guessing took about 9h), but we
failed to derive a recurrence by creative telescoping, which would prove our guess.

\begin{conjecture}
  If $(a_n)$ denotes the sequence \seqnum{A172671} then for
  $\tilde{a}_n := \frac{n!^3}{(3n)!}a_n$ we have
  \begin{align*}
    & 3 (n+3) (n+4)^3 (3784 n^4+32164 n^3+100749 n^2+137862 n+69678) \tilde{a}_{n+4} \\
    & {}-(n+3) (3799136 n^7+72183584 n^6+579689880 n^5+2548427912 n^4 \\
    & \qquad {}+6617561702 n^3+10141503096 n^2+8487349821 n+2991586122) \tilde{a}_{n+3} \\
    & {}-3 (10844944 n^8+222321352 n^7+1973930222 n^6+9916013134 n^5 \\
    & \qquad {}+30831383530 n^4+60768378830 n^3+74160044251 n^2\\
    & \qquad {}+51243135187 n+15352797306) \tilde{a}_{n+2} \\
    & {}+(n+2) (29681696 n^7+504588832 n^6+3602458816 n^5+14001842392 n^4 \\
    & \qquad {}+32010306742 n^3+43078657918 n^2+31639900193 n\\
    & \qquad{} +9799573455) \tilde{a}_{n+1} \\
    & {}+15435 (n+1)^3 (n+2) (3784 n^4+47300 n^3+219945 n^2\\
    & \qquad {}+450988 n+344237) \tilde{a}_n = 0.
  \end{align*}
\end{conjecture}

Although we cannot prove that the conjectured recurrences for \seqnum{A172671} and \seqnum{A172572}
are correct, it follows from the sum expressions that some recurrences for these sequences
must exist.

\begin{theorem}
  \seqnum{A172671} and \seqnum{A172572} are D-finite. 
\end{theorem}

\subsection{\texorpdfstring{Sequence \seqnum{A188818}}{A188818}}\label{sec:A188818}

This sequence counts the number of $n\times n$ binary arrays without the
pattern $01$ diagonally or antidiagonally. The OEIS lists 32 terms, from which
the LLL-based guesser finds a recurrence of order~5 and degree~10. With LA-based
guessing one needs at least 55 terms to find this recurrence. Although it is not
obvious at first glance, also here lattice paths turn out to be the key to the
solution.

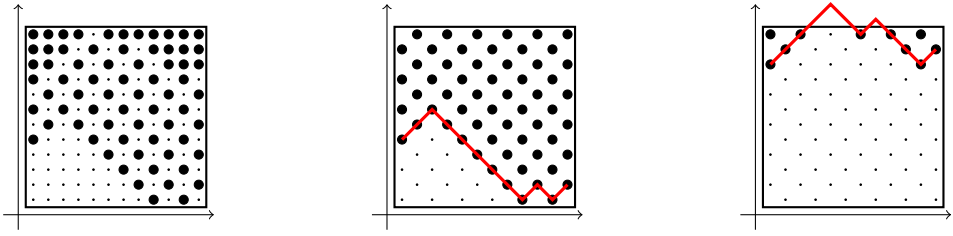
\begin{figure}
\begin{center}
\begin{tikzpicture}[scale=0.2]
  \draw[->] (-1,0) -- (13,0);
  \draw[->] (0,-1) -- (0,14);
  \draw[thick] (0.5,0.5) rectangle (12.5,12.5);
  \foreach \x in {1,...,12} {
    \foreach \y in {1,...,12} {\draw[fill,radius=0.05] (\x,\y) circle;}
  }
  \draw[fill,radius=0.3] (1,12) circle;
  \foreach \i in {1,2,3} {\draw[fill,radius=0.3] (\i,9+\i) circle;}
  \draw[fill,radius=0.3] (7,12) circle;
  \foreach \i in {1,2,3} {\draw[fill,radius=0.3] (8+\i,13-\i) circle;}
  \draw[fill,radius=0.3] (11,12) circle;
  \draw[fill,radius=0.3] (12,11) circle;
  \draw[fill,radius=0.3] (1,5) circle;
  \draw[fill,radius=0.3] (2,6) circle;
  \draw[fill,radius=0.3] (1,7) circle;
  \foreach \i in {1,...,9} {\draw[fill,radius=0.3] (\i,10-\i) circle;}
  \foreach \i in {1,...,11} {\draw[fill,radius=0.3] (\i,12-\i) circle;}
  \foreach \i in {1,...,11} {\draw[fill,radius=0.3] (1+\i,13-\i) circle;}
  \foreach \i in {1,...,9} {\draw[fill,radius=0.3] (3+\i,13-\i) circle;}
  \foreach \i in {1,...,7} {\draw[fill,radius=0.3] (5+\i,13-\i) circle;}
  \foreach \i in {1,...,5} {\draw[fill,radius=0.3] (7+\i,13-\i) circle;}
  \foreach \i in {1,...,3} {\draw[fill,radius=0.3] (9+\i,13-\i) circle;}
  \draw[fill,radius=0.3] (12,12) circle;
\end{tikzpicture}
\hfil
\begin{tikzpicture}[scale=0.2]
  \draw[->] (-1,0) -- (13,0);
  \draw[->] (0,-1) -- (0,14);
  \draw[thick] (0.5,0.5) rectangle (12.5,12.5);
  \foreach \x in {1,3,...,11} {
    \foreach \y in {1,3,...,11} {\draw[fill,radius=0.05] (\x,\y) circle;}
    \foreach \y in {2,4,...,12} {\draw[fill,radius=0.05] (1+\x,\y) circle;}
  }
  \draw[fill,radius=0.3] (1,5) circle;
  \draw[fill,radius=0.3] (2,6) circle;
  \draw[fill,radius=0.3] (1,7) circle;
  \foreach \i in {1,...,9} {\draw[fill,radius=0.3] (\i,10-\i) circle;}
  \foreach \i in {1,...,11} {\draw[fill,radius=0.3] (\i,12-\i) circle;}
  \foreach \i in {1,...,11} {\draw[fill,radius=0.3] (1+\i,13-\i) circle;}
  \foreach \i in {1,...,9} {\draw[fill,radius=0.3] (3+\i,13-\i) circle;}
  \foreach \i in {1,...,7} {\draw[fill,radius=0.3] (5+\i,13-\i) circle;}
  \foreach \i in {1,...,5} {\draw[fill,radius=0.3] (7+\i,13-\i) circle;}
  \foreach \i in {1,...,3} {\draw[fill,radius=0.3] (9+\i,13-\i) circle;}
  \draw[fill,radius=0.3] (12,12) circle;
  \draw[very thick,red,-] (1,5) -- (3,7) -- (9,1) -- (10,2) -- (11,1) -- (12,2);
\end{tikzpicture}
\hfil
\begin{tikzpicture}[scale=0.2]
  \draw[->] (-1,0) -- (13,0);
  \draw[->] (0,-1) -- (0,14);
  \draw[thick] (0.5,0.5) rectangle (12.5,12.5);
  \foreach \x in {1,3,...,11} {
    \foreach \y in {2,4,...,12} {\draw[fill,radius=0.05] (\x,\y) circle;}
    \foreach \y in {1,3,...,11} {\draw[fill,radius=0.05] (1+\x,\y) circle;}
  }
  \draw[fill,radius=0.3] (1,12) circle;
  \foreach \i in {1,2,3} {\draw[fill,radius=0.3] (\i,9+\i) circle;}
  \draw[fill,radius=0.3] (7,12) circle;
  \foreach \i in {1,2,3} {\draw[fill,radius=0.3] (8+\i,13-\i) circle;}
  \draw[fill,radius=0.3] (11,12) circle;
  \draw[fill,radius=0.3] (12,11) circle;
  \draw[very thick,red,-] (1,10) -- (5,14) -- (7,12) -- (8,13) -- (11,10) -- (12,11);
\end{tikzpicture}
\end{center}
\caption{A particular $12\times12$ array for \seqnum{A188818} (left), where dots and
  bullets represent 0's and 1's, respectively. The middle (resp., right) image shows the
  even (resp., odd) positions, where 0's and 1's are separated by a 
  generalized Dyck path.}
\label{fig:A188818}
\end{figure}

The fact that the forbidden patterns are considered along (anti-)diagonals
allows us to decompose the problem. The even positions in the array, i.e.,
positions $(x,y)$ with $x+y$ even, and the odd positions can be filled with
0's and 1's independent of each other. Hence, the $n$th sequence term, $a_n$,
can be written as
\[
  a_n = e_n \cdot o_n,
\]
where $e_n$ and $o_n$ count the number of admissible $\{0,1\}$-arrangements on
the even and odd positions in the $n\times n$ array, respectively.
See Fig.~\ref{fig:A188818}, where a particular solution (left) is decomposed into
an even part (middle) and an odd part (right).

If we focus only on positions of the same parity, then we see that the array contains
a region with 1's on the top, and at the bottom a region with 0's. Both regions are
separated by a path that starts somewhere on the left border, ends somewhere on the
right border, and uses steps $(1,1)$ and $(1,-1)$ (see Fig.~\ref{fig:A188818}).
Let $D\bigl((a,b)\to(c,d)\mathrel{\big|}R\bigr)$ denote the number of such generalized
Dyck paths that start at $(a,b)$, end at $(c,d)$, and satisfy certain restrictions~$R$.

In our setting, we certainly have the restriction $y\geq1$ to avoid that the
path leaves the $n\times n$ array through its bottom side. For the upper side,
we have to allow the path to leave the square a little bit, in order to enable
0's to appear in the top row, but the path must not go above $n+2$ (see the
right part of Fig.~\ref{fig:A188818}). For example, to compute $e_n$ for
odd~$n$, we add up
\[
  D\bigl((1,y_1)\to(n,y_2)\mathrel{\big|}1\leq y\leq n+2\bigr)
\]
for $y_1=1,3,\dots,n+2$ and $y_2=1,3,\dots,n+2$, 
and similarly for even~$n$, and analogously for~$o_n$. Since the paths are restricted to a
rectangle which is higher than wide, no path could ever violate the lower and
upper restriction at the same time. Hence we can rewrite
\begin{alignat*}1
  &D\bigl((1,y_1)\to(n,y_2)\mathrel{\big|}1\leq y\leq n+2\bigr)\\
  & \hbox{\quad\quad} =
  \begin{cases}
    D\bigl((1,y_1)\to(n,y_2)\mathrel{\big|}y\geq1\bigr), & y_1+y_2\leq n+1; \\
    D\bigl((1,y_1)\to(n,y_2)\mathrel{\big|}y\leq n+2\bigr), & \text{otherwise}.
  \end{cases}
\end{alignat*}
By mirroring horizontally, we obtain
\[
  D\bigl((1,y_1)\to(n,y_2)\mathrel{\big|}y\leq n+2\bigr) =
  D\bigl((1,n+3-y_1)\to(n,n+3-y_2)\mathrel{\big|}y\geq1\bigr).
\]
By combining equal cases and by substituting $y_1\to2k+1$ and $y_2\to n+2-2\ell$,
we can write
\begin{alignat*}1
  e_n &= \sum_{k=0}^{\lfloor\frac{n+1}{2}\rfloor} \biggl(
  D\bigl((1,2k+1)\to(n,n+2-2k)\mathrel{\big|}y\geq1\bigr)\\
  &\qquad{}+
  2\cdot\!\sum_{\ell=k+1}^{\lfloor\frac{n+1}{2}\rfloor}
  D\bigl((1,2k+1)\to(n,n+2-2\ell)\mathrel{\big|}y\geq1\bigr)\biggr)
\end{alignat*}
and a similar expression for $o_n$. The generalized Dyck paths are counted by
a difference of binomial coefficients,
\begin{alignat*}1
  &D\bigl((x_1,y_1)\to(x_2,y_2)\mathrel{\big|}y\geq1\bigr)\\
  &\hbox{\quad\quad} =\binom{x_2-x_1}{\frac12(x_2-x_1+y_2-y_1)} - \binom{x_2-x_1}{\frac12(x_2-x_1+y_2+y_1)},
\end{alignat*}
which follows from \cite[Theorem~10.3.1]{krattenthaler15}, after the Dyck paths
have been translated to simple lattice paths via the substitution
$(x,y)\to((x+y-2)/2,(x-y)/2)$ for even points, and $(x,y)\to((x+y-1)/2,(x-y+1)/2)$
in the case of odd points. We insert this closed form expression for~$D$, simplify
a bit, and end up with the following expressions for $e_n$ and $o_n$:
\begin{align*}
  e_n &= 2^{n-2} + 2\cdot\sum_{k=0}^{\lfloor\frac{n+1}{2}\rfloor} \sum_{\ell=k+1}^{\lfloor\frac{n+1}{2}\rfloor}
  \left(\binom{n-1}{n-k-\ell} - \binom{n-1}{n+k-\ell+1}\right), \\
  o_n &= 2^{n-2} + 2\cdot\sum_{k=0}^{\lfloor\frac{n}{2}\rfloor} \sum_{\ell=k+1}^{\lfloor\frac{n}{2}\rfloor}
  \left(\binom{n-1}{n-k-\ell-1} - \binom{n-1}{n+k-\ell+1}\right).
\end{align*}
Creative telescoping delivers provably correct recurrences for $e_n$ and $o_n$,
which by closure properties can be combined to a recurrence for $a_n$. Since the
corresponding order-42 operator is a left multiple of our guessed order-5 operator,
we have established the correctness of our guess.

\begin{theorem}
  \seqnum{A188818} is D-finite and satisfies a recurrence of order~5 and degree~10.
\end{theorem}

\subsection{\texorpdfstring{Sequence \seqnum{A306322}}{A306322}}\label{sec:A306322}

Here we count $n\times n$ integer matrices $((m_{i,j}))_{i,j=1}^n$ with $m_{1,1}=0$ and $m_{n,n}=2$,
and all rows, columns, and falling diagonals weakly monotonic without jumps of~2. An example for
$n=7$ is given by
\begin{center}
  \def\a{0}\def\b{1}\def\c{2}
  \begin{tikzpicture}[scale=.6,yscale=.8]
    \draw[xshift=-.5cm,yshift=-.5cm] (0,0)grid(7,7);
    \draw(0,6) node {\a} ++(1,0) node {\a} ++(1,0) node {\a} ++(1,0) node {\a} ++(1,0) node {\a} ++(1,0) node {\a} ++(1,0) node {\a};
    \draw(0,5) node {\a} ++(1,0) node {\a} ++(1,0) node {\a} ++(1,0) node {\a} ++(1,0) node {\a} ++(1,0) node {\a} ++(1,0) node {\b};
    \draw(0,4) node {\a} ++(1,0) node {\a} ++(1,0) node {\a} ++(1,0) node {\b} ++(1,0) node {\b} ++(1,0) node {\b} ++(1,0) node {\b};
    \draw(0,3) node {\a} ++(1,0) node {\a} ++(1,0) node {\a} ++(1,0) node {\b} ++(1,0) node {\b} ++(1,0) node {\b} ++(1,0) node {\b};
    \draw(0,2) node {\a} ++(1,0) node {\a} ++(1,0) node {\a} ++(1,0) node {\b} ++(1,0) node {\c} ++(1,0) node {\c} ++(1,0) node {\c};
    \draw(0,1) node {\a} ++(1,0) node {\a} ++(1,0) node {\b} ++(1,0) node {\b} ++(1,0) node {\c} ++(1,0) node {\c} ++(1,0) node {\c};
    \draw(0,0) node {\a} ++(1,0) node {\b} ++(1,0) node {\b} ++(1,0) node {\c} ++(1,0) node {\c} ++(1,0) node {\c} ++(1,0) node {\c};
  \end{tikzpicture}.
\end{center}
The key to recognizing this sequence as D-finite is hidden in the OEIS-entry of the bivariate sequence \seqnum{A323846},
which is defined analogously for a $k\times n$ matrix. It is remarked there that
the problem goes back to Knuth~\cite{knuth19} and that the labels $0$, $1$, and $2$
divide the matrix into three connected regions, so that counting the number of matrices is equivalent to counting
pairs of non-intersecting lattice walks from the lower left to the upper right corner. It is well-known that such
pairs of lattice walks are counted by the Narayana numbers, but this is not quite the final answer. Two adjustments
need to be made: (1)~there must be at least one $0$ in the top-left corner and at least one $2$ in the bottom-right
corner, and (2)~the walk-pairs are not required to start and end in the corners.

\begin{figure}
\begin{center}
  \leavevmode\llap{a)\quad}
  \begin{tikzpicture}[scale=.4]
    \draw(0,0) rectangle (5,5) (2.5,0) node[below] {$n$} (0,2.5) node[left] {$n$};
    \draw(0,2) rectangle (3,5) (3,3.5) node[right] {$j$} (1.5,2) node[below] {$i$};
    \draw(0,2)--(1,3.5)--(2,4)--(2.3,4.25)--(2.5,4.5)--(3,5);
    \draw(0,2)--(.5,2.5)--(2,3)--(2.8,4)--(3,5);
  \end{tikzpicture}\hfil
  \llap{b)\quad}
  \begin{tikzpicture}[scale=.4]
    \draw(0,0) rectangle (5,5) (2.5,0) node[below] {$n$} (0,2.5) node[left] {$n$};
    \begin{scope}[xshift=2cm,yshift=-2cm]
      \draw(0,2) rectangle (3,5) (0,3.5) node[left] {$j$} (1.5,5) node[above] {$i$};
      \draw(0,2)--(1,3.5)--(2,4)--(2.3,4.25)--(2.5,4.5)--(3,5);
      \draw(0,2)--(.5,2.5)--(2,3)--(2.8,4)--(3,5);
    \end{scope}
    \draw(5,2.5) node[right] {\hphantom{$j$}};
  \end{tikzpicture}

  \leavevmode\llap{c)\quad}
  \begin{tikzpicture}[scale=.4]
    \draw(0,0) rectangle (5,5) (2.5,0) node[below] {$n$} (0,2.5) node[left] {$n$};
    \begin{scope}[xshift=1cm,yshift=-3.333cm,yscale=1.6667]
      \draw(0,2) rectangle (3,5) (0,5) node[above] {$i$} (3,5) node[above] {$j$};
      \draw(0,2)--(1,3.5)--(2,4)--(2.3,4.25)--(2.5,4.5)--(3,5);
      \draw(0,2)--(.5,2.5)--(2,3)--(2.8,4)--(3,5);
    \end{scope}
  \end{tikzpicture}\hfil
  \llap{d)\quad}
  \begin{tikzpicture}[scale=.4]
    \draw(0,0) rectangle (5,5) (2.5,0) node[below] {$n$} (0,2.5) node[left] {$n$};
    \begin{scope}[rotate=90,xshift=1cm,yshift=3.333cm,yscale=-1.6667]
      \draw(0,2) rectangle (3,5) (0,5) node[right] {$i$} (3,5) node[right] {$j$};
      \draw(0,2)--(1,3.5)--(2,4)--(2.3,4.25)--(2.5,4.5)--(3,5);
      \draw(0,2)--(.5,2.5)--(2,3)--(2.8,4)--(3,5);
    \end{scope}
  \end{tikzpicture}  
  
\end{center}
\caption{Case distinction used in the analysis of \seqnum{A306322}.}\label{fig:A306322}
\end{figure}
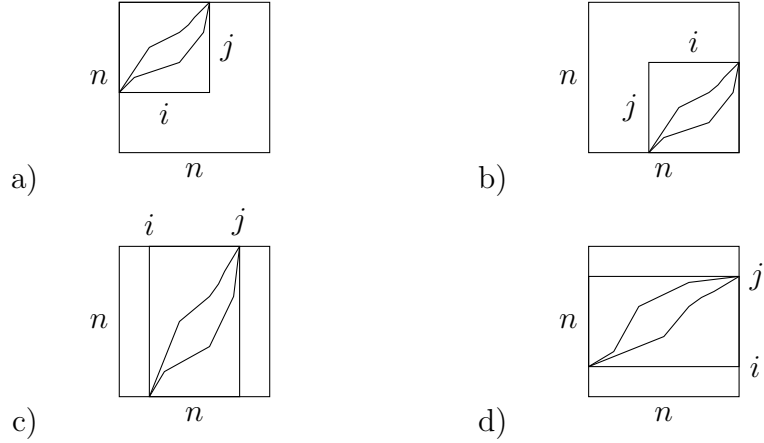

We know that $N_{i,j}=\frac1{i+j-1}\binom{i+j-1}i\binom{i+j-1}{i-1}$ is the number
of non-intersecting walk-pairs in an $i\times j$ board, and that $\binom{i+j}{i}$ is
the total number of walks in such a board. Therefore
\[
  \sum_{i,j=1}^n\biggl(N_{i,j} - \binom{i+j-2}{i-1}\biggr) - \binom{2n}n + 1
\]
is the number of walk-pairs of the form shown in Fig.~\ref{fig:A306322}~a),
excluding the walk-pairs where the upper walk passes through the upper-left corner
(accounted for by the term $\binom{i+j-2}{i-1}$) as well as the walk-pairs where the
lower walk passes through the lower-right corner (accounted for by the term $\binom{2n}n$;
the $1$ accounts for the doubly excluded walk-pair where the upper walk passes through
the top-left corner and the lower walk through the lower-right corner).

The same expression also counts the walks of the form shown in Fig.~\ref{fig:A306322}~b), with the analogous exceptions removed.
Taking both cases together, we count the cases $i=j=n$ twice, so we altogether only have
\[
2\biggl(\sum_{i,j=1}^n\biggl(N_{i,j} - \binom{i+j-2}{i-1}\biggr) - \binom{2n}n + 1\biggr) - \biggl(N_{n,n} - 2\binom{2n}n + 1\biggr)
\]
such walk-pairs. We also have to take into account walk-pairs of the form shown in Fig.~\ref{fig:A306322} c) and~d).
In both cases, their number is $\sum_{i=1}^{n-1}\sum_{j=i+1}^{n-1} N_{j-i,n}$, where the boundaries of the sum are chosen so that
we do not count anything that was already counted before. In conclusion, we find the expression
\begin{alignat*}1
&2\biggl(\sum_{i,j=1}^n\biggl(N_{i,j} - \binom{i+j-2}{i-1}\biggr) - \binom{2n}n + 1 + \sum_{i=1}^{n-1}\sum_{j=i+1}^{n-1} N_{j-i,n}\biggr)\\
&- \biggl(N_{n,n} - 2\binom{2n}n + 1\biggr)
\end{alignat*}
for the $n$th term of \seqnum{A306322}. Clearly this is D-finite. 

Using
\[
\sum_{i,j=1}^n\binom{i+j-2}{i-1}=\binom{2n}n-1
\]
and
\[
\sum_{i=1}^{n-1}\sum_{j=i+1}^{n-1} N_{j-i,n}=\sum_{j=1}^n\sum_{i=1}^{n-j-1} N_{j,n},
\]
the expression can be simplified to
\begin{alignat*}1
  &2\sum_{j=1}^n\sum_{i=1}^nN_{i,j} + 2\sum_{j=1}^{n-1}(n-j-1) N_{j,n} - 2\binom{2n}n - N_{n,n} + 3\\
  &=2\sum_{j=1}^n\sum_{i=1}^nN_{i,j} + 2\sum_{j=1}^{n}(n-j-1) N_{j,n} - 2\binom{2n}n + N_{n,n} + 3\\
  &=2\sum_{j=1}^n\biggl(\sum_{i=1}^nN_{i,j} + (n-j-1)N_{j,n}\biggr) - 2\binom{2n}n + N_{n,n} + 3.
\end{alignat*}

The \texttt{HolonomicFunctions.m} package~\cite{koutschan10c} effortlessly obtains
for this expression an operator of order~12 and degree~87 that
contains the guessed recurrence as right factor. 

\begin{theorem}
  \seqnum{A306322} is D-finite and satisfies a recurrence of order~4 and degree~14.
\end{theorem}

\section{Further examples}\label{sec:pa}

\subsection{\texorpdfstring{Sequence \seqnum{A195806} and \seqnum{A216940}}{A195806 and A216940}}
\label{sec:A195806}

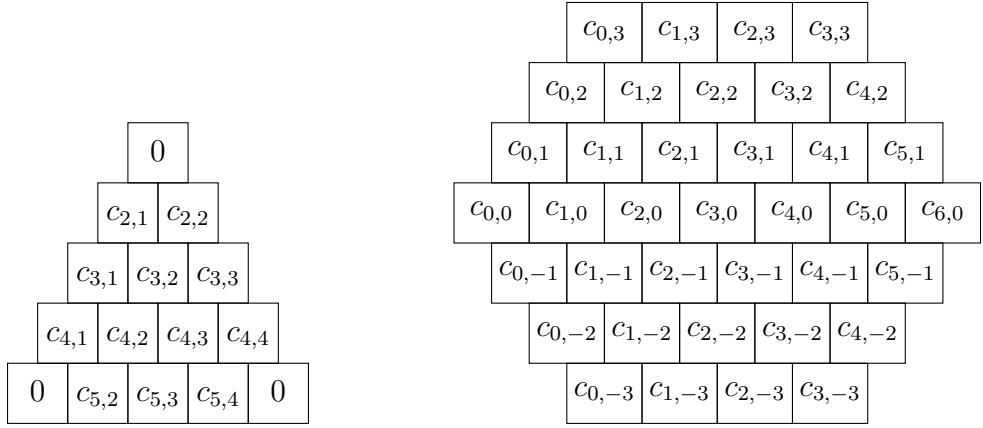
\begin{figure}
\begin{center}
    \begin{tikzpicture}[scale=.8]
      \foreach \i/\j in {0/0, -1/-.5, -1/.5, -2/-1, -2/0, -2/1, -3/-1.5, -3/-.5, -3/.5, -3/1.5, -4/-2, -4/-1, -4/0, -4/1, -4/2}
      \draw (\j,\i) rectangle +(1,1);
      \begin{scope}[xshift=.5cm,yshift=.5cm]
        \draw (0,0) node {$\mathstrut 0$} (-2,-4) node {$\mathstrut 0$} (2,-4) node {$\mathstrut 0$};
        \draw (-.5,-1) node {$\mathstrut c_{2,1}$} (.5,-1) node {$\mathstrut c_{2,2}$};
        \draw (-1,-2) node {$\mathstrut c_{3,1}$} (0,-2) node {$\mathstrut c_{3,2}$} (1,-2) node {$\mathstrut c_{3,3}$};
        \draw (-1.5,-3) node {$\mathstrut c_{4,1}$} (-.5,-3) node {$\mathstrut c_{4,2}$} (.5,-3) node {$\mathstrut c_{4,3}$} (1.5,-3) node {$\mathstrut c_{4,4}$};
        \draw (-1,-4) node {$\mathstrut c_{5,2}$} (0,-4) node {$\mathstrut c_{5,3}$} (1,-4) node {$\mathstrut c_{5,4}$};
      \end{scope}
    \end{tikzpicture}
    \hfil
  \begin{tikzpicture}[yscale=.8]
    \foreach \x in {0,...,6} \draw (\x,0) rectangle +(1,1) (\x+.5,.5) node {$c_{\x,0}$};
    \foreach \x in {0,...,5}
    \draw (\x+.5,1) rectangle +(1,1) (\x+1,1.5) node {$c_{\x,1}$}
          (\x+.5,-1) rectangle +(1,1) (\x+1,-.5) node {$c_{\x,-1}$};
    \foreach \x in {0,...,4}
    \draw (\x+1,2) rectangle +(1,1) (\x+1.5,2.5) node {$c_{\x,2}$}
          (\x+1,-2) rectangle +(1,1) (\x+1.5,-1.5) node {$c_{\x,-2}$};
    \foreach \x in {0,...,3}
    \draw (\x+1.5,3) rectangle +(1,1) (\x+2,3.5) node {$c_{\x,3}$}
          (\x+1.5,-3) rectangle +(1,1) (\x+2,-2.5) node {$c_{\x,-3}$};
  \end{tikzpicture}
\end{center} 
\caption{Illustrations of the arrays appearing in the definitions of \seqnum{A195806} (left)
  and \seqnum{A216940} (right), respectively.}\label{fig:omega}
\end{figure}

For the sequence \seqnum{A195806}, we count triangular arrays of size~5 whose entries are chosen from $\{0,\dots,n\}$
in such a way that all rows and diagonals having the same length have the same sums, and with $0$
assigned to the corners (cf. Fig.~\ref{fig:omega}).
The specification of this sequence can be easily translated into a system of linear inequalities. The $n$th term
of the sequence is precisely the number of integer solutions of the following equations and inequalities: 
\begin{alignat*}1
  &0\leq c_{i,j}\leq n\quad\text{for all $i,j$},\\
  &c_{2,1}+c_{2,2}=c_{4,1}+c_{5,2}=c_{5,4}+c_{4,4},\\
  &c_{3,1}+c_{3,2}+c_{3,3}=c_{3,1}+c_{4,2}+c_{5,3}=c_{3,3}+c_{4,3}+c_{5,3},\\
  &c_{4,1}+c_{4,2}+c_{4,3}+c_{4,4}=c_{2,2}+c_{3,2}+c_{4,2}+c_{5,2}=c_{2,1}+c_{3,2}+c_{4,3}+c_{5,4},\\
  &c_{2,1}+c_{3,1}+c_{4,1}=c_{5,2}+c_{5,3}+c_{5,4}=c_{2,2}+c_{3,3}+c_{4,4}.
\end{alignat*}
Partition analysis provides theory and algorithms for dealing with such systems. From the theory, which has its
roots in the early 20th century~\cite{macmahon15}, it follows immediately that the sequence \seqnum{A195806} is
a quasipolynomial.
In particular, it must be D-finite. With the associated algorithms~\cite{andrews01}, it is possible to compute the
quasipolynomial explicitly, at least in principle. With the implementations we had available, the computation
did not complete in a reasonable amount of time. However, the recurrence found by our LLL-based guesser suggests
the following expression.

\begin{conjecture} If $(a_n)$ denotes the sequence \seqnum{A195806}, then
\begin{alignat*}1
  a_n&= \frac1{1296}\bigl(130n^6 + 1560n^5 + 8125n^4 + 23400n^3\bigr)\\
  &\quad{}+ \frac1{1296}\left\{\begin{array}{ll}
40788 n^2+42768 n+20736, & \text{if $n\equiv0\ (\mathrm{mod}\ 6)$;}\\
40692 n^2+42128 n+20045, & \text{if $n\equiv1\ (\mathrm{mod}\ 6)$;}\\
40788 n^2+42256 n+19712, & \text{if $n\equiv2\ (\mathrm{mod}\ 6)$;}\\
40788 n^2+42768 n+20493, & \text{if $n\equiv3\ (\mathrm{mod}\ 6)$;}\\
40692 n^2+42128 n+20288, & \text{if $n\equiv4\ (\mathrm{mod}\ 6)$;}\\
40788 n^2+42256 n+19496, & \text{if $n\equiv5\ (\mathrm{mod}\ 6)$.}
\end{array}\right.
\end{alignat*}
\end{conjecture}

The sequence \seqnum{A216940} is quite similar. Here we count hexagonal arrays of size~4 filled with
elements of $\{0,\dots,n\}$ in such a way that the entries are nondecreasing towards east, south west,
and south east (cf. Fig.~\ref{fig:omega}).
Again, the specification can be easily translated into a system of linear inequalities, so it follows
immediately that the sequence is a quasipolynomial and in particular D-finite.
Again, we were not able to derive an expression by a rigorous computation based on partition analysis,
but we had no trouble to find a solution from our guessed recurrence.
In fact, it appears that the result is not only a quasipolynomial but a polynomial.

\begin{conjecture} If $(a_n)$ denotes the sequence \seqnum{A216940}, then
\begin{alignat*}1
a_n&=
(n+1)^{\overline{13}} (n+6)^{\overline3} (n+7) (74384146n^{20}+10413780440 n^{19}\\
&+694580474022 n^{18}+29345762188932 n^{17}+880856790135603 n^{16}\\
&+19969728998781072 n^{15}+354853893929158096n^{14}\\
&+5062226797216352960 n^{13}+58900361433618244860 n^{12}\\
&+564694034848365996336 n^{11}+4487557575514810132362n^{10}\\
&+29630015361661371290844 n^9+162382123713323392711687 n^8\\
&+735273283907306553706472 n^7+2726904840964417033376520n^6\\
&+8166353315859794719296864 n^5+19314394347459920710102704 n^4\\
&+34829846371335010335540480 n^3+45137854540680193956153600 n^2\\
&+37557333457279933473792000 n+15118483615575730790400000)\\
&/221424599279703105635713957232640000000,
\end{alignat*}
where we use the raising factorial notation $x^{\overline k}=x(x+1)\cdots(x+k-1)$. 
\end{conjecture}
Incidentally, the degree of this polynomial matches the number of terms that were given in the OEIS.

Although we were not able to prove that our guessed recurrences are correct, partition analysis
implies that the sequences are quasi-polynomials, and are therefore D-finite. 

\begin{theorem}
  \seqnum{A195806} and \seqnum{A216940} are D-finite. 
\end{theorem}

\subsection{\texorpdfstring{Sequence \seqnum{A194478}}{A194478}}\label{sec:A194478}

For this sequence, we consider a triangular grid of varying size, and the question is how
many ways there are to arrange 6 indistinguishable points on it in such a way that no
three points are in the same row or diagonal.

For $n=5$, an example for such an arrangement is
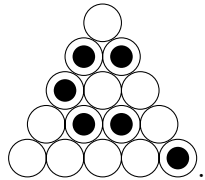
\begin{center}
  \begin{tikzpicture}[scale=.5]
    \draw (0,0) circle (.5cm);
    \draw (-.5,-.9) circle (.5cm) (.5,-.9) circle (.5cm);
    \draw (-1,-1.8) circle(.5cm) (0,-1.8) circle(.5cm) (1,-1.8) circle (.5cm);
    \draw (-1.5,-2.7) circle(.5cm) (-.5,-2.7) circle(.5cm) (.5,-2.7) circle(.5cm) (1.5,-2.7) circle(.5cm);
    \draw (-2,-3.6) circle(.5cm) (-1,-3.6) circle(.5cm) (0,-3.6) circle(.5cm) (1,-3.6) circle(.5cm) (2,-3.6) circle(.5cm);
    \foreach \x/\y in {-1/-1.8,-.5/-2.7,.5/-.9,.5/-2.7,-.5/-.9,2/-3.6} \fill (\x,\y) circle(.3cm);
  \end{tikzpicture}.
\end{center}
The $n$th term of the sequence \seqnum{A194478} is the number of such arrangements for a triangle of size~$n$.
The sequence is the 6th column of the bivariate sequence \seqnum{A194480}, where guessed polynomial
expressions are given for the first five columns. According to our guessed recurrence, the 6th
column is not a polynomial but the quasipolynomial
\begin{alignat*}1
  &\frac{1}{256} (-1)^n (2 n-7)(n^2-7 n+13)+ \frac{1}{322560}(7 n^{12}+42n^{11}-945 n^{10}\\
  &\quad+1274 n^9+26089 n^8-128810 n^7+175693 n^6+205366n^5-810796 n^4\\
  &\quad+601328 n^3+354172 n^2-582180 n+114660).
\end{alignat*}
Note that the degree and the leading coefficient of this quasipolynomial are consistent with the
degrees and leading coefficients of the guessed polynomials for the earlier columns.

We prove the correctness of the above expression using the principle of inclusion/exclusion.
Let $a^{(i)}(n,k)$ denote the number of ways to select $k$ places from a triangle of size~$n$
in such a way that at least $i$ lines (rows or diagonals) contain three or more selected places,
counted with multiplicities. 
The number of interest is then
\[
  a(n,k) = a^{(0)}(n,k) - a^{(1)}(n,k) + a^{(2)}(n,k) - a^{(3)}(n,k) \pm \cdots.
\]
We have $a^{(0)}(n,k) = \binom{\binom{n+1}2}{k}=\frac1{2^kk!}n^{2k}+\O(n^{2k-1})$.
Next, for each $i\in\{1,\dots,n\}$ there are altogether three lines of length~$i$,
and for each of them there are $\binom ij$ ways to select $j$ positions on it,
and $\binom{\binom{n+1}2-i}{k-j}$ ways to choose $k-j$ positions in the remaining
triangle. Thus
\[
a^{(1)}(n,k) = 3\sum_{j=3}^6\sum_{i=1}^n\binom ij\binom{\binom{n+1}2-i}{k-j}.
\]

In order to count how many ways there are to have at least two lines with three
selected positions, we distinguish three cases. In case~1, the two lines have the
same orientation (i.e., they are parallel). Restricting now for simplicity to $k=6$,
we then have to select three places on each line, which can be done in 
$3\sum_{i=1}^n\sum_{j=1}^{i-1}\binom{i}{3}\binom{j}3$
many ways. In case~2, the two lines have different orientation (i.e., they
are not parallel), but they have no intersection point. This happens when the lengths
of the lines add up to at most~$n$, so there are
$3\sum_{i=1}^n\sum_{j=1}^{n-i}\binom{i}{3}\binom{j}3$
such arrangements. In case~3, we have two lines that do intersect. This case has two
subcases, depending on whether the intersection point is selected or not. If it is
selected, only five positions are required to be on the two lines and the sixth
position can be selected arbitrarily from the remaining triangle
(either on none of the lines or on the first line or on the second line).
This makes 
\begin{alignat*}1
  &3\sum_{i=1}^n\sum_{j=n-i+1}^n\biggl(\binom{i-1}{2}\binom{j-1}{2}\binom{\binom{n+1}2-i-j+1}{1}\\
  &\qquad+\binom{i-1}{3}\binom{j-1}{2}+\binom{i-1}{2}\binom{j-1}{3}\biggr)
\end{alignat*}
possibilities in this case. Finally, there are
\[
3\sum_{i=1}^n\sum_{j=n-i+1}^n\binom{i-1}{3}\binom{j-1}{3}
\]
arrangements where the two lines intersect but the intersection point is not among
the selected positions. Altogether,
\begin{alignat*}1
a^{(2)}(n,6)&=6\sum_{i=1}^n\sum_{j=1}^{n-i}\binom{i}{3}\binom{j}3+3\sum_{i=1}^n\sum_{j=n-i+1}^n\binom{i-1}{3}\binom{j-1}{3}\\
&+3\sum_{i=1}^n\sum_{j=n-i+1}^n\biggl(\binom{i-1}{2}\binom{j-1}{2}\binom{\binom{n+1}2-i-j+1}{1}\\
&\qquad{}+\binom{i-1}{3}\binom{j-1}{2}+\binom{i-1}{2}\binom{j-1}{3}\biggr).
\end{alignat*}
If there are three lines with at least three selected positions, then, as there are
altogether only six selected positions, three of them must belong to two lines. In
particular, the three lines must have pairwise distinct orientation, and
they must not intersect in the same position. Then each line contains two intersection
points and one additional selected position. This makes
\[
a^{(3)}(n,6)
= \sum_{i=3}^n\sum_{j=n-i+1}^n (i-2)(j-2)\biggl(
     \sum_{\ell=n-\min(i,j)+1}^{2n-(i+j)}(\ell-2) + \sum_{\ell=2n+2-(i+j)}^n(\ell-2)
     \biggr).
\]
Since $a^{(m)}(n,6)=0$ for $m\geq4$, we have
\[
 a_n = a(n,6) = a^{(0)}(n,6) - a^{(1)}(n,6) + a^{(2)}(n,6) - a^{(3)}(n,6),
\]
and while this is an expression of intimidating length, it must be observed that all the lower
arguments of the binomials are explicit integers, so the sums are in fact just polynomial
sums. It is the $\min(i,j)$ appearing in one of the summation boundaries in the expression
for $a^{(3)}(n,6)$ which is responsible for the fact that the $a_n$ is not a polynomial but
only a quasipolynomial.

\begin{theorem}
  If $(a_n)$ denotes the sequence \seqnum{A194478}, then
  \begin{alignat*}1
    &a_n=\frac{1}{256} (-1)^n (2 n-7)(n^2-7 n+13)+ \frac{1}{322560}(7 n^{12}+42n^{11}-945 n^{10}\\
    &\quad+1274 n^9+26089 n^8-128810 n^7+175693 n^6+205366n^5-810796 n^4\\
    &\quad+601328 n^3+354172 n^2-582180 n+114660).
  \end{alignat*}
\end{theorem}

\section{Conjectures}\label{sec:open}

\subsection{\texorpdfstring{Sequence \seqnum{A215570}}{A215570}}\label{sec:A215570}

Now we want to count the number of permutations of $n$ copies of $\{1,\dots,5\}$,
as in Sect.~\ref{sec:A177317}, but with a more complicated condition:
every partial sum is at most the same partial sum averaged over all permutations.
In other words, the $k$th partial sum of the permutation must not exceed~$3k$,
because the average $(1+2+3+4+5)/5$ is equal to~$3$.

The OEIS displays a dynamic programming code for enumerating such
permutations. For fixed integer~$n$, let $b_{v,w,x,y,z}$ denote the number
of permutations of length~$5n-v-\dots-z$ with $n-v$ 1's, $n-w$ 2's,
etc., and satisfying the partial-sums condition. This means that still
$v$ 1's, $w$ 2's, etc.\ have to be appended, to turn them into
permutations of the desired form. From the values of $v,\dots,z$ one can deduce
which numbers are allowed to be appended next, yielding a set of rules
to compute the five-dimensional sequence $b_{v,w,x,y,z}$ recursively.  For
example, $b_{3,2,0,1,4}$ means that one has to put the total amount of
$3\cdot1+2\cdot2+1\cdot4+4\cdot5=31$ onto the remaining $3+2+0+1+4=10$ places,
which means that we can exceed the average of~$3$ by at most $31-3\cdot10=1$.
Hence, the number~$5$ must be excluded, as well as the number~$3$ (because the
third index is equal to~$0$), and we get
\[
  b_{3,2,0,1,4} = b_{2,2,0,1,4} + b_{3,1,0,1,4} + b_{3,2,0,0,4}.
\]
Finally, then $n$th sequence term $a_n$ is computed by applying this rule
recursively to $b_{n,n,n,n,n}$ until the termination condition $b_{0,0,0,0,0}=1$ is
reached. This procedure runs reasonably fast, by caching intermediate
values, but has high memory consumption. Computing the first 51 terms, 
approximately the amount of data given in the OEIS, took about 2.5 hours
and required 60\,GB of memory. Obviously, more terms could only be obtained
at a significant computational cost.

The above transition rules can equivalently be encoded
in a transfer matrix. The states are given by the possible margins one
has to remember when appending new numbers. In the worst case, where the
permutation starts with all $1$'s and $2$'s, the margin can go up to $3n$,
and thus we get a $(3n+1)\times(3n+1)$ matrix. As in Sect.~\ref{sec:tm}, 
we have to introduce catalytic variables $x_i$ for
recording how often the number~$i$ has occurred.
This way we can obtain the values $a_n$ with less memory consumption,
but the timing is much longer (21~hours for the first 51 terms).
The transfer matrix is a Toeplitz matrix of bandwidth~2,
\[
  \setlength{\arraycolsep}{2pt}
  M = \left(\begin{array}{ccccc}
    x_3 & x_4 & 1 & 0 & \cdots \\[-4pt]
    x_2 & x_3 & x_4 & 1 & \ddots \\[-4pt]
    x_1 & x_2 & x_3 & x_4 & \ddots \\[-4pt]
    0 & x_1 & x_2 & x_3 & \ddots \\[-4pt]
    \vdots & \ddots & \ddots & \ddots & \ddots
  \end{array}\right).
\]

Can we now conclude that \seqnum{A215570} is D-finite and derive a
corresponding recurrence? No, unfortunately not. Like already seen
in the example of Sect.~\ref{sec:A250556}, the matrix here does not have a
fixed dimension. For fixed~$n$, the same $(3n+1)\times(3n+1)$ matrix can be used
to compute all the values $a_0,\dots,a_n$, but not beyond.
Hence, we leave our guessed recurrence
as a conjecture and invite the reader to prove that it is correct.
We note that the recurrence becomes simpler when we consider a related
sequence, that differs from the original one by a hypergeometric factor.

\begin{conjecture}
  If $(a_n)$ denotes the sequence \seqnum{A215570} then for the
  auxiliary sequence $\tilde{a}_n :=  \frac{n!^3(n+1)!^2}{(5n)!}a_n$ we have
  \begin{align*}
    & 3 (3 n+8) (3 n+10) (65 n^3+398 n^2+781 n+496) \tilde{a}_{n+3} \\
    & {}-4 (910 n^5+11032 n^4+52047 n^3 +119686 n^2+134365 n+58980) \tilde{a}_{n+2} \\
    & {}+(2015 n^5+24428 n^4+114387 n^3 +258294 n^2+281088 n+118368) \tilde{a}_{n+1} \\
    & {}-2 (n+1) (n+2) (65 n^3+593 n^2+1772 n+1740) \tilde{a}_n = 0.
  \end{align*}
\end{conjecture}

The OEIS also has related entries where $n$ copies of $\{1,\dots,m\}$ are
considered, the above discussion referring to the special case $m=5$.
For $m=1,2,3$, the resulting sequences are D-finite (in fact, hypergeometric).
For $m=4$ (\seqnum{A215562}), there are 134 known terms, but surprisingly they
are not sufficient for guessing a recurrence, not even with LLL-based
guessing.  The relevant average in this case is $\frac14(1+2+3+4)=\frac52$,
which means that the transfer matrix needs to be twice as big as expected,
because the margins have to be considered in steps of~$\frac12$. Equivalently,
one can use two different transfer matrices, which are multiplied in turn,
depending on whether an even or odd position is filled.  This somewhat
explains why the case $m=4$ is harder than~$m=5$. In addition, the sequence
terms have much fewer small integer factors, and thus it seems unlikely that
transforming the sequence with a hypergeometric factor would simplify the
guessing problem.

It remains an open problem to find a provably correct recurrence equation satisfied by the sequence \seqnum{A215562}.

\subsection{\texorpdfstring{Sequence \seqnum{A339987}}{A339987}}\label{sec:A339987}

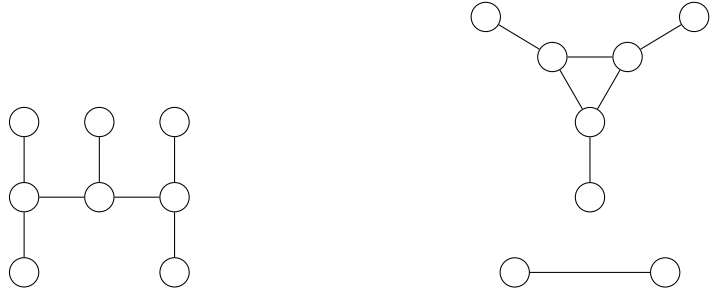
\begin{figure}
\begin{center}
  \begin{tikzpicture}[draw,circle]
    \node[draw,circle](0) at (0,0) {};
    \node[draw,circle](1) at (0,1) {};
    \node[draw,circle](2) at (0,-1) {};
    \node[draw,circle](3) at (1,0) {};
    \node[draw,circle](4) at (1,1) {};
    \node[draw,circle](5) at (2,0) {};
    \node[draw,circle](6) at (2,1) {};
    \node[draw,circle](7) at (2,-1) {};
    \draw (0)--(1) (0)--(2) (0)--(3) (3)--(4) (3)--(5) (5)--(6) (5)--(7);
  \end{tikzpicture}
  \hfil
  \begin{tikzpicture}
    \node[draw,circle] (0) at (0,0) {};
    \node[draw,circle] (1) at (-.5,0.866025) {};
    \node[draw,circle] (2) at (.5,0.866025) {};
    \begin{scope}[yshift=-1cm,scale=2.77]
      \node[draw,circle] (3) at (0,0) {};
      \node[draw,circle] (4) at (-.5,0.866025) {};
      \node[draw,circle] (5) at (.5,0.866025) {};
    \end{scope}
    \node[draw,circle] (6) at (-1,-2) {};
    \node[draw,circle] (7) at (1,-2) {};
    \draw (6)--(7) (0)--(1)--(2)--(0) (0)--(3) (1)--(4) (2)--(5);
  \end{tikzpicture}
\end{center}
\caption{Two graphs with 8 vertices used for illustrating the definition of \seqnum{A339987}.}\label{fig:graphs}
\end{figure}

This sequence is defined as the number of labeled graphs on $2n$ vertices that share the same degree
sequence as any unrooted binary tree on $2n$ vertices. This means that $n-1$ vertices must have degree~3
and the remaining $n+1$ vertices must have degree~1.
For example, for $n=4$, there are only the following two unlabeled graphs with this property (Fig.~\ref{fig:graphs}).
The graph shown in Fig.~\ref{fig:graphs} on the left can be labeled in $8\cdot\binom72\cdot5\cdot\binom42=5040$ ways,
and the graph shown on the right (consisting of two connected components) can be labeled in $\binom83\cdot5\cdot4\cdot3=3360$
ways. Consequently, we have $a_4=8400$. 

We found a recurrence for the sequence $(a_n)$ of order~5 with polynomial coefficients of degree~10.
Its polynomial coefficients contain several low-degree factors, which provides some evidence in favor
of the recurrence. It also suggests to write $a_n = \frac1{n+1}(\frac52)^{\overline{n-2}}\tilde a_n$ for
some other auxiliary sequence~$(\tilde a_n)$. The recurrence for $(a_n)$ translates into a recurrence
for $(\tilde a_n)$ which also has order~5 but polynomial coefficients of lower degree.

\begin{conjecture} If $(a_n)$ denotes the sequence \seqnum{A339987} and we set
  $\tilde a_n=a_n/(\frac1{n+1}(\frac52)^{\overline{n-2}})$, then
  \begin{alignat*}1
    &1024 (n+2) (328 n^3+3300 n^2+10844 n+11589)\tilde a_n\\
    &-128 (2624 n^4+30664 n^3+129460 n^2+232328 n+148119)\tilde a_{n+1}\\
    &-128 (2952 n^5+40852 n^4+219308 n^3+569267 n^2+712135 n+341634)\tilde a_{n+2}\\
    &+32 (3936 n^5+55672 n^4+306380 n^3+818282 n^2+1057879 n+527520)\tilde a_{n+3}\\
    &-4 (2624 n^5+42472 n^4+264028 n^3+786236 n^2+1117119 n+601452)\tilde a_{n+4}\\
    &+3 (n+4) (328 n^3+2316 n^2+5228 n+3717)\tilde a_{n+5}=0.
  \end{alignat*}
\end{conjecture}
Observe that the cubic factor in the coefficient of $\tilde a_n$ can be obtained from the cubic factor in
the coefficient of $\tilde a_{n+5}$ by setting $n$ to $n+1$. This is another property that we would not
expect to encounter on a wrongly guessed recurrence.

According to Maple, the linear operator corresponding to the recurrence for $(\tilde a_n)$ is irreducible.
Experimentally, we find the asymptotic expansion
\[
  \tilde a_n\sim c\,n!(\frac{32}3)^n\Bigl(1 + \frac7{256}n^{-1}-\frac{55023}{131072}n^{-2}-\frac{13563843}{33554432}n^{-3}+\O(n^{-4})\Bigr)
\]
for a constant
\begin{alignat*}1
  c=0.&7269505475849839203724738433453909726988076\_\\
      &\_083835242155944045267221957561211243532139\dots
\end{alignat*}

\subsection{\texorpdfstring{Sequence \seqnum{A269021}}{A269021}}\label{sec:A269021}

Sequences related to pattern avoiding permutations have been intensively studied~\cite{vatter15}.
In this context, some sequences are known to be D-finite, others are known not
to be D-finite, and there are some for which the status is open. A prominent
example is the sequence of 1324-avoiders (\seqnum{A061552}), of which only 50
terms are known~\cite{conway15}. We have not found any recurrence candidate based on
these terms, and recent empirical arguments~\cite{conway18} suggest that the
sequence is more likely not D-finite than D-finite.

It is known~\cite{gessel90,bostan20} that for every fixed~$k$, the number of permutations of
length $n$ avoiding the pattern $123\cdots k$ is D-finite as a sequence in~$n$.
However, this result has no immediate implications on sequences we obtain when
$n$ and $k$ are coupled. For example, the sequence \seqnum{A269021} is defined as the
number of permutations of length $2n$ containing the pattern $123\cdots n$. (Obviously,
counting permutations that do contain a given pattern is as easy or difficult as
counting permutations that do not.) From the 42 terms given in the OEIS, we were able
to detect a recurrence of order~4 and degree~21. This recurrence has the hypergeometric
term $(n-1)(2n)!$ among its solutions.

\begin{conjecture}
  If $(a_n)$ denotes the sequence \seqnum{A269021}, and we set $\tilde a_n=a_n/(2n)!^2$, then
  \begin{alignat*}1
    &(-64 n^{10}-1968 n^9-26156 n^8-198469 n^7-952323 n^6-3012795 n^5\\
    &\qquad -6333869 n^4-8663374 n^3-7264534 n^2-3266000 n-549760) \tilde a_n\\
    &+(64 n^{13}+2672 n^{12}+49788 n^{11}+545913 n^{10}+3917758 n^9+19359535 n^8\\
    &\quad +67385886 n^7+165789363 n^6+284054698 n^5+325846005 n^4\\
    &\quad+229526554 n^3+78563984 n^2-487964 n-5543040) \tilde a_{n+1}\\
    &+(-512 n^{15}-21568 n^{14}-419248 n^{13}-4969164 n^{12}-39928763 n^{11}\\
    &\quad -228837227 n^{10}-959068672 n^9-2966908118 n^8-6753094929 n^7\\
    &\quad -11118771121 n^6-12741784568 n^5-9313604242 n^4-3271711596 n^3\\
    &\quad +562569136 n^2+946158512 n+250467360) \tilde a_{n+2}\\
    &+2 (n+3) (512 n^{16}+26752 n^{15}+624800 n^{14}+8677944 n^{13}+80260596 n^{12}\\
    &\quad +523718876 n^{11}+2488583381 n^{10}+8747566435 n^9+22820793074 n^8\\
    &\quad +43766004538 n^7+60004107039 n^6+55047935941 n^5+27672902302 n^4\\
    &\quad-778719870 n^3-10812498240 n^2-6360099840 n-1300242000) \tilde a_{n+3}\\
    &-12 (n+4)^3 (n+3) (2 n+7)^2 (3 n+8) (3 n+10) (64 n^{10}+1328 n^9+11324 n^8\\
    &\quad +52389 n^7+143536 n^6+233810 n^5+204716 n^4+48699 n^3-68928 n^2\\
    &\quad -61278 n-15900) \tilde a_{n+4}=0.
  \end{alignat*}  
\end{conjecture}

\subsection{\texorpdfstring{Sequence \seqnum{A181198} and \seqnum{A181199}}{A181198 and A181199}}
\label{sec:A181198}

We find a recurrence of order~2 and degree~9 for the sequence \seqnum{A181198} based on the 27 terms that were given in the database,
but in this instance we realized that this is not too impressive a discovery because it is easy to generate enough further terms that
LA-based guessing can find the recurrence.

The sequence is defined as the number of $(4\times n)$-matrices filled with the numbers $1,\dots,4n$ in such a way
that all rows, columns, diagonals, and antidiagonals (downwards) are increasing. An example for $n=4$ is
\begin{center}
  \begin{tabular}{|c|c|c|c|}
    \hline
    1 & 2 & 3 & 4 \\\hline
    5 & 6 & 7 & 8 \\\hline
    9 & 10 & 12 & 14 \\\hline
    11 & 13 & 15 & 16 \\\hline
  \end{tabular}\;.
\end{center}
Here is a way to count such matrices efficiently. Assume that we fill the
$4\times n$ array with the numbers $1,\dots,4n$ in that order. Then at each
intermediate step the filled cells must form a Young diagram (so that the
condition of increasing values row- and column-wise is satisfied), plus the
extra condition that these Young diagrams must not have two rows of equal
length, unless these have length~$n$ (this is to ensure the
antidiagonally-increasing condition). We need not care about the
diagonally-increasing condition, as this one is automatically implied by the
first two. We want to count the number of ways how to transform the empty
Young diagram $(0,0,0,0)$ into the rectangle $(n,n,n,n)$, according to the
above rules. Let us encode the situation as a formal sum of terms $c\cdot
x_{s,t,u,v}$, which transport the information that there have been $c$ ways to
produce the Young diagram corresponding to the partition $(s,t,u,v)$. Then
adding a box to the diagram corresponds to the application of the rule
\begin{align*}
  x_{s,t,u,v} \to{}
  & [s<n]\cdot x_{s+1,t,u,v} + {} \\
  & [t<s-1 \lor t=n-1]\cdot x_{s,t+1,u,v} + {} \\
  & [u<t-1 \lor u=n-1]\cdot x_{s,t,u+1,v} + {} \\
  & [v<u-1 \lor v=n-1]\cdot x_{s,t,u,v+1},
\end{align*}
where $[P]$ denotes the Iverson bracket. For example,
\[
  x_{5,3,2,0} \to x_{6,3,2,0} + x_{5,4,2,0} + x_{5,3,2,1},
\]
assuming that $n>5$. In order to compute $a_n$, we start with the expression
$x_{0,0,0,0}$, then apply the above rule $4n$ times (i.e., in each of the $4n$
rounds we apply it to each occurrence of $x_{s,t,u,v}$), and we will end up
with the expression $a_nx_{n,n,n,n}$.
An implementation in Mathematica takes about 25 minutes to get the first 100
terms of the sequence. This is more than enough to find the recurrence with
LA-based guessing.

The guessed recurrence suggests a closed form expression.

\begin{conjecture} If $(a_n)$ denotes the sequence \seqnum{A181198}, then for $n>1$ we have
\begin{alignat*}1
a_n&=\frac{(-64)^n (n-1) (-\frac{1}{2})^{\overline{2 n}} (\frac{1}{2})^{\overline{n}}}{4(3 n)!}\\
   &\times\biggl(-1 + 3 \sum_{k=2}^{n-1}
  \frac{(-4)^k (7 k^2-1) }{(k-1) k (k+1)^2 (2 k-1)^2 (2 k+1)^3} \binom{3 k}{2 k} \binom{k+\frac{1}{2}}{k}
  \biggr)
\end{alignat*}
\end{conjecture}

As an example for guessing with little data, the related sequence \seqnum{A181199}
is more interesting. It is defined in the same way as \seqnum{A181198}, just
with $(5\times n)$-matrices instead of $(4\times n)$-matrices. The OEIS listed
only 26 terms, which was not enough for the LLL-based guesser to find any recurrence.
However, by the procedure outlined above, we were able to produce 60 terms,
and this is more than enough for the LLL-based guesser to detect a convincing recurrence
of order~3 and degree~24. The LA-based guesser would need more than 100 terms
to find this recurrence, and with our implementation it takes more than 14 hours
to produce them.

According to Maple, the operator corresponding to the recurrence admits a factorization
as a product of three operators of order~1. This factorization suggests again an
explicit expression for the sequence.

\begin{conjecture}\allowdisplaybreaks
If $(a_n)$ denotes the sequence \seqnum{A181199}, then
\[
 a_n = 1 - \frac{27}4\sum_{k=1}^{n-1}(-1)^k u(k)\frac{(5k)!}{(3k)!k!^2}\sum_{i=1}^{k-1}(-1)^iv(i)\frac{(3i)!}{i!^3}
\]
where
\begin{alignat*}1
  u(k) &= 8\,\bigl(25216 k^8+9888 k^7-14496 k^6+11208 k^5+23832 k^4+7383 k^3 \\
  &\qquad -1522 k^2-939 k-90\bigr)
  \big/\bigl((2k-1) (4 k-1) (3 k+1)^{\overline{3}} (4 k+1)^{\overline{4}}\bigr), \\
  v(i)&=\bigl((3 i+1) (3 i+2) (4 i+3) (137855872 i^{11}+860969696 i^{10}\\
  &\qquad+2047036856 i^9+2032587274 i^8-24192441 i^7-1894061166 i^6\\
  &\qquad-1671661480 i^5-524330624 i^4+36004789 i^3+62751860 i^2\\
  &\qquad+13865604 i+927360)\bigr)\\
  &\quad\big/\bigl((i+1)^2 (i+2)^2 (2 i-1) (2 i+1) (2 i+3) (25216 i^8+9888 i^7-14496 i^6\\
  &\qquad+11208 i^5+23832 i^4+7383 i^3-1522 i^2-939 i-90) (25216 i^8\\
  &\qquad+211616 i^7+760768 i^6+1543976 i^5+1973632 i^4+1683047 i^3\\
  &\qquad+971955 i^2+353502 i+60480)\bigr).
\end{alignat*}
\end{conjecture}

\subsection{\texorpdfstring{Sequence \seqnum{A181280}}{A181280}}\label{sec:A181280}

For every $n\in\set N$, the $n$th term of this sequence is defined as the number of
matrices $M\in\set Z_2^{4\times n}$ with the following properties:
\begin{itemize}
\item The rows of $M$, read as bit strings, are lexicographically strictly increasing.
\item The rows of $MM^\top\in\set Z_2^{4\times 4}$, read as bit strings, are lexicographically strictly decreasing. 
\end{itemize}
The OEIS entry contains the following example for $n=5$:
\[
M=\begin{pmatrix}
0&1&0&1&1\\
1&0&0&0&0\\
1&1&0&0&1\\
1&1&1&1&0
\end{pmatrix}\quad\Rightarrow\quad
MM^\top=\begin{pmatrix}
1&0&0&0\\
0&1&1&1\\
0&1&1&0\\
0&1&0&0
\end{pmatrix}.
\]
The recurrence we found for this sequence suggests the following closed form expression for
the sequence.

\begin{conjecture} If $(a_n)$ denotes the sequence \seqnum{A181280}, then for $n\geq4$ we have
  \begin{alignat*}1
    a_n &=
    \tfrac{1}{3} 2^{2 n-11} (6 n^2-219 n+820)-\tfrac{1}{9} 2^{n-5} (3 n+32)
    -\tfrac{113}{3} (-1)^n 2^{3 n-14}\\
    &\quad+2^{4 n-9}-\tfrac{1}{3} (-1)^n 2^{2n-11} (13 n-164)+\tfrac{1}{9} 2^{3 n-14} (288 n-3473).
  \end{alignat*}
\end{conjecture}

\subsection{\texorpdfstring{Sequence \seqnum{A253217}}{A253217}}\label{sec:A253217}

This sequence has a somewhat complicated definition. Its $n$th term is the number
of ways to fill an $n\times n$ array with nonnegative integers in such a way that
the following conditions are satisfied:
\begin{itemize}
\item The entry at position $(1,1)$ is $0$ and the entry at position $(n,n)$ is $n-3$.
\item The entry at each position $(i,j)$ is either equal to or one more than the entries
  at positions $(i-1,j)$, $(i,j-1)$, and $(i-1,j-1)$. 
\item The entry at each position $(i,j)$ belongs to $\{\max(i,j)-2,\max(i,j)-1,\max(i,j)\}$
\end{itemize}
An example for $n=8$ is the array
\begin{center}
  \begin{tabular}{|c|c|c|c|c|c|c|c|}\hline
    0 & 1 & 1 & 2 & 3 & 4 & 5 & 5 \\\hline
    1 & 1 & 2 & 2 & 3 & 4 & 5 & 5 \\\hline
    2 & 2 & 2 & 2 & 3 & 4 & 5 & 5 \\\hline
    2 & 2 & 3 & 3 & 3 & 4 & 5 & 5 \\\hline
    3 & 3 & 3 & 3 & 3 & 4 & 5 & 5 \\\hline
    4 & 4 & 4 & 4 & 4 & 4 & 5 & 5 \\\hline
    4 & 5 & 5 & 5 & 5 & 5 & 5 & 5 \\\hline
    5 & 5 & 5 & 5 & 5 & 5 & 5 & 5 \\\hline
  \end{tabular}.
\end{center}

The sequence \seqnum{A253217} is the diagonal of the bivariate sequence~\seqnum{A253223}, where the
counting problem is considered more generally for rectangular arrays. In the entry
for this bivariate sequence, it is conjectured that all rows and columns are
ultimately quadratic polynomials.

\begin{conjecture}\allowdisplaybreaks If $(a_n)$ denotes the sequence $\seqnum{A253217}$, then
  \begin{alignat*}1
    &32 (n+1) (2 n+1)^2 (1575 n^6+21285 n^5+117954 n^4+343020 n^3\\
    &\quad +551943 n^2+465785 n+161046) a_n\\
    &-8 (121275 n^9+1933470 n^8+13267683 n^7+51280818 n^6+122556360 n^5\\
    &\quad +186866686 n^4+180574335 n^3+105734340 n^2+33718283 n\\
    &\quad +4443102) a_{n+1}\\
    &+2 (294525 n^9+4763070 n^8+33170868 n^7+130145646 n^6+315713355 n^5\\
    &\quad +488415476 n^4+478464380 n^3+283626704 n^2+91378536 n\\
    &\quad +12137328) a_{n+2}\\
    &+ (294525 n^9+4668570 n^8+31877118 n^7+122735586 n^6+292620525 n^5\\
    &\quad +445804136 n^4+431097970 n^3+252913504 n^2+80866406 n\\
    &\quad +10688508) a_{n+3}\\
    &- (121275 n^9+1961820 n^8+13655808 n^7+53503836 n^6+129484209 n^5\\
    &\quad +199650088 n^4+194784258 n^3+114948300 n^2+36871922 n\\
    &\quad +4877748) a_{n+4}\\
    &+2 (n+3)^2 (2 n+7) (1575 n^6+11835 n^5+35154 n^4+52554 n^3+41382 n^2\\
    &\quad +16118 n+2428) a_{n+5}=0.
  \end{alignat*}
\end{conjecture}

The conjectured recurrence has the exact solutions $1$, $(-2)^n$, and~$4^n$ and two further solutions
whose asymptotic expansions have the dominant terms $(\tfrac14)^nn^{-1/2}$ and $16^nn^{-1}$, respectively.
For the generating function $\sum_{n=0}^\infty a_nx^n$, we found a convincing differential equation
of order~4 and degree~15; the corresponding differential operator $L$ can be factored as a product
$L=L_1L_2L_3$ where $L_1$ has order~2 and $L_2,L_3$ both have order~1.

\subsection{\texorpdfstring{Sequence \seqnum{A098926}}{A098926}}\label{sec:A098926}

The $n$th term of this sequence is defined as the permanent of the $(n+2)\times(n+2)$ matrix
where the entry at position $(i,j)$ is zero if $(i,j)$ belongs to the path that starts
at $(1,1)$ and alternatingly moves two steps to the right and two steps down. All other
entries are~$1$. For example, the 8th term of the sequence is the permanent of the
matrix
\[
\begin{pmatrix}
  0 & 0 & 0 & 1 & 1 & 1 & 1 & 1 & 1 & 1 \\
  1 & 1 & 0 & 1 & 1 & 1 & 1 & 1 & 1 & 1 \\
  1 & 1 & 0 & 0 & 0 & 1 & 1 & 1 & 1 & 1 \\
  1 & 1 & 1 & 1 & 0 & 1 & 1 & 1 & 1 & 1 \\
  1 & 1 & 1 & 1 & 0 & 0 & 0 & 1 & 1 & 1 \\
  1 & 1 & 1 & 1 & 1 & 1 & 0 & 1 & 1 & 1 \\
  1 & 1 & 1 & 1 & 1 & 1 & 0 & 0 & 0 & 1 \\
  1 & 1 & 1 & 1 & 1 & 1 & 1 & 1 & 0 & 1 \\
  1 & 1 & 1 & 1 & 1 & 1 & 1 & 1 & 0 & 0 \\
  1 & 1 & 1 & 1 & 1 & 1 & 1 & 1 & 1 & 1 
\end{pmatrix}.
\]

\begin{conjecture} If $(a_n)$ denotes the sequence \seqnum{A098926}, then
  \begin{alignat*}1
    &n (n+1)(3 n^5+95 n^4+1113 n^3+5983 n^2+14907 n+14025) a_n\\
    &-(n+1)(13 n^4+388 n^3+3717 n^2+13424 n+16865) a_{n+1}\\
    &-(9 n^7+294 n^6+3677 n^5+22722 n^4+76591 n^3+146304 n^2\\
    &\quad +157554 n+81720) a_{n+2}\\
    &-(n^5-103 n^4-2125 n^3-14395 n^2-38283 n-32845) a_{n+3}\\
    &+(9 n^7+318 n^6+4409 n^5+30672 n^4+113879 n^3+219268 n^2\\
    &\quad +186788 n+35600) a_{n+4}\\
    &+(17 n^5+445 n^4+4253 n^3+17161 n^2+24893 n+1765) a_{n+5}\\
    &-(3 n^7+122 n^6+2039 n^5+18038 n^4+90333 n^3+252920 n^2\\
    &\quad +364438 n+211080) a_{n+6}    \\
    &-(3 n^5+83 n^4+833 n^3+3663 n^2+6967 n+4465) a_{n+7}\\
    &+(3 n^5+80 n^4+763 n^3+3184 n^2+5915 n+4080) a_{n+8}=0.
  \end{alignat*}
\end{conjecture}

Besides the recurrence stated above, we also found a
convincing differential equation of order~3 and degree~19 for which the corresponding
differential operator $L$ can be written as a product of three operators of order~1.
This means that $L$ can be solved in terms of d'Alembertian solutions. In fact, it appears
that the generating function $\sum_{n=0}^\infty a_nx^n$ can be written as
\begin{alignat*}1
&c\frac{x^2-x-2}{x(x-1)}\exp\Bigl(\frac{x+1}{x(x-1)}\Bigr)\\
&\quad\times\int^x r(y)\exp\Bigl(\frac{-2y^2-2}{y(y-1)(y+1)}\Bigr)
\int^y s(z)\exp\Bigl(\frac{z-1}{z(z+1)}\Bigr)dz\,dy
\end{alignat*}
with
\begin{alignat*}1
r(y)&=\frac{y^5-3y^4+2y^3-2y^2-y+1}{y(y+1)^4(y-2)^2},\\
s(z)&=\frac{z^2(z-2)(z^8-2z^7-12z^6+28z^5-10z^4-22z^3+4z^2+4z+1)}{(z-1)^2(z^5-3z^4+2z^3-2z^2-z+1)^2},
\end{alignat*}
and for a suitably chosen constant $c$ and suitably chosen constants of integration.

\subsection{\texorpdfstring{Sequence \seqnum{A164735}}{A164735}}\label{sec:A164735}

The Kaprekar map \seqnum{A151949} is defined as follows. Given an integer~$n$,
read it as a string of (decimal) digits, without any leading zeros. Sort the
characters once in decreasing order and once in increasing order. Read these
two strings again as integers and subtract the smaller from the larger. The
resulting number is the image of~$n$.

For example, $n=64308654$ is mapped to
\[
86654430 - 03445668=83208762
\]
by this process, $n=83208762$ is mapped to $88763220 - 02236788=86526432$, and
$n=86526432$ is mapped to $86654322 - 22345668=64308654$. It turns out that
we have a cycle of length three: $64308654\to83208762\to86526432\to64308654$.

The sequence of interest is not the Kaprekar map itself, but a sequence that
counts the number of such cycles: The $n$th term of \seqnum{A164735} is defined
as the number of cycles of length three among all the integers with $n$ decimal
digits. For $n=8$, there is no other cycle besides the one stated above, so
the $8$th term of \seqnum{A164735} is~$1$.

The LLL-based guesser detected a recurrence of order~15 and degree~4 from the 70 terms
listed in the OEIS. The recurrence can be solved in terms of quasipolynomials,
leading to the following conjecture:
\begin{conjecture}\label{conj:kaprekar} If $(a_n)$ denotes the sequence \seqnum{A164735}, then
  for all $n\geq3$
\[
  a_{18k+i} = \frac{1}{40}\!
  \begin{cases}
    3 (243 k^5+405 k^4+35 k^3+395 k^2-318 k+40), & i=0; \\
    k (729 k^4-405 k^3-615 k^2+225 k+106), & i=1; \\
    729 k^5+1620 k^4+735 k^3+1320 k^2-684 k+40, & i=2; \\
    k (729 k^4-705 k^2+136), & i=3; \\
    3 k (243 k^4+675 k^3+515 k^2+565 k-118), & i=4; \\
    k (729 k^4+405 k^3-615 k^2-225 k+106), & i=5; \\
    3 k (243 k^4+810 k^3+845 k^2+790 k+32), & i=6; \\
    3 k (k+1) (243 k^3+27 k^2-142 k+12), & i=7; \\
    729 k^5+2835 k^4+3705 k^3+3405 k^2+726 k+40, & i=8; \\
    3 k (k+1) (243 k^3+162 k^2-127 k-18), & i=9; \\
    729 k^5+3240 k^4+5055 k^3+4860 k^2+1636 k+160, & i=10; \\
    3 k (k+1) (243 k^3+297 k^2-52 k-48), & i=11; \\
    729 k^5+3645 k^4+6585 k^3+6795 k^2+2926 k+400, & i=12; \\
    3 k (k+1) (243 k^3+432 k^2+83 k-58), & i=13; \\
    729 k^5+4050 k^4+8295 k^3+9270 k^2+4696 k+800, & i=14; \\
    3 k (k+1) (243 k^3+567 k^2+278 k-28), & i=15; \\
    3 (k+3) (243 k^4+756 k^3+1127 k^2+734 k+160), & i=16; \\
    3 k (k+1) (243 k^3+702 k^2+533 k+62), & i=17.
  \end{cases}
  \]
\end{conjecture}
We are able to identify two patterns that yield numbers in Kaprekar 3-cycles.
Using word notation, e.g., $1^4=1111$, the first one reads
\[
  X_{m,a,b,c,d,e} := 9^e8^m7^d6^m5^c4^m3^b2^m1^a09^m8^{a+1}7^m6^b5^m4^c3^m2^d1^m0^{e-1}1
\]
$(m,a,b\geq0,\,c,d,e\geq1)$.
A direct calculation shows that the Kaprekar map sends $X_{m,a,b,c,d,e}$ to
$X_{m,c-1,b,d,a+1,e}$, which is sent to $X_{m,d-1,b,a+1,c,e}$, which finally is sent
back to $X_{m,a,b,c,d,e}$. Hence we have a 3-cycle, except if $a+1=c=d$ in which
case we run into a 1-cycle. The number $X_{m,a,b,c,d,e}$ has $2(a+b+c+d+e+1)+9m$
digits, and therefore $m$ is forced to have the same parity as~$n$. For
example, for odd~$n$ the number of 3-cycles is given by
\begin{alignat*}1
  \tfrac13\bigl|\bigl\{
  X_{2\ell+1,a,b,c,d,e} \mathrel{\big|}\;
  &0\leq\ell\leq\bigl\lfloor\tfrac{n-17}{18}\bigr\rfloor,\,
  a,b\geq0,\,c,d,e\geq1,\\
  &a+b+c+d+e=\tfrac{n-18\ell-11}{2},\,\neg(a+1=c=d)
  \bigr\}\bigr|,
\end{alignat*}
which indeed yields the polynomial expressions displayed above, and which
explains the period 18 of the conjectured quasi-polynomial. For even~$n$
we can write down a similar expression, but this is not enough. There is
a second pattern,
\[
  Y_{a,b,c} := 65^c43^b1^a08^{a+1}6^b54^{c+1}
  \quad (a,c\geq0,\,b\geq1),
\]
which produces only integers with an even number of digits. Again, it
is not difficult to see that each $Y_{a,b,c}$ gives rise to a 3-cycle under
the Kaprekar map (but note that the other two members of each cycle are not of
the form $Y_{a',b',c'}$). The only 3-cycle of 8-digit numbers mentioned above
is generated by $Y_{0,1,0}$.
For even~$n$, the two patterns give the following number of 3-cycles:
\begin{alignat*}1
  &\tfrac13\bigl|\bigl\{
  X_{2\ell,a,b,c,d,e} \mathrel{\big|}
  0\leq\ell\leq\bigl\lfloor\tfrac{n-8}{18}\bigr\rfloor,\,
  a,b\geq0,\,c,d,e\geq1,\\
  & \phantom{\tfrac13\bigl|\bigl\{ X_{2\ell,a,b,c,d,e} \mathrel{\big|}{}}
  a+b+c+d+e=\tfrac{n-18\ell-2}{2},\,
  \neg(a+1=c=d)\bigr\}\bigr|\\
  &{}+ \bigl|\bigl\{
  Y_{a,b,c} \mathrel{\big|} a,c\geq0,\,b\geq1,\,a+b+c=\tfrac{n-6}{2}
  \bigr\}\bigr|.
\end{alignat*}
As before, this produces the other half of the quasi-polynomial expression
that was conjectured above. While these considerations shed some light on the
occurrence of a complicated-looking quasi-polynomial of period~18, they do not
prove anything. In view of the number-theoretic flavor of the construction, we
could well imagine that the conjectured expression is only valid until a
certain (possibly large) limiting index~$n$ and then breaks down, because
further patterns for members of 3-cycles may appear. Among all the conjectures
stated in this paper, Conj.~\ref{conj:kaprekar} is the one in which we believe
least.

\section{Acknowledgments}
We thank Neil Sloane, Vaclav Kotesovec, Christian Krattenthaler,
Doron Zeilberger, Paul Zimmermann, Stefan Gerhold, and Alin Bostan for their
interest in our work and for enlightening discussions, Erich Kaltofen for
making us aware that the recurrence for \seqnum{A172671} can be simplified
by removing a hypergeometric factor, and Carsten Schneider for the suggestion
to look for d'Alembertian solutions.
Both authors acknowledge support of the Austrian FWF grant I6130-N.
MK moreover acknowledges support of the Austrian FWF grant P31571-N32.

	\bigskip
	\hrule
	\bigskip
	
	\noindent 
	2020 \emph{Mathematics Subject Classification}:~Primary 05A15.
        Secondary 68W30, 33F10.
	
	\noindent 
	\emph{Keywords}:~guessing, recurrence equations, D-finiteness, computer algebra.
	
	\bigskip
	\hrule
	\bigskip
	
	\noindent 
	Concerned with sequences
\seqnum{A039836},
\seqnum{A061552},
\seqnum{A098926},
\seqnum{A151949},
\seqnum{A164735},
\seqnum{A172572},
\seqnum{A172671},
\seqnum{A177317},
\seqnum{A181198},
\seqnum{A181199},
\seqnum{A181280},
\seqnum{A187990},
\seqnum{A188818},
\seqnum{A194478},
\seqnum{A194480},
\seqnum{A195806},
\seqnum{A199250},
\seqnum{A215562},
\seqnum{A215570},
\seqnum{A216940},
\seqnum{A237684},
\seqnum{A250556},
\seqnum{A253217},
\seqnum{A253223},
\seqnum{A264946},
\seqnum{A264947},
\seqnum{A265234},
\seqnum{A269021},
\seqnum{A306322},
\seqnum{A323846},
\seqnum{A331562}, and
\seqnum{A339987}.

\bigskip
\hrule
\bigskip

\vspace*{+.1in}
\noindent
Published in 
\htmladdnormallink{Journal of Integer Sequences}{https://cs.uwaterloo.ca/journals/JIS/},
April 24 2023.



\end{document}